%% file: bare_jrnl_new_sample4.tex
\documentclass[10pt,journal,compsoc]{IEEEtran}

\usepackage{amsmath,amsfonts}
\usepackage{algorithmic}
\usepackage{algorithm}
\usepackage{array}
\usepackage[caption=false,font=normalsize,labelfont=sf,textfont=sf]{subfig}
\usepackage{textcomp}
\usepackage{stfloats}
\usepackage{url}
\usepackage{verbatim}
\usepackage{graphicx}
\usepackage{cite}
\usepackage{scalerel}
\usepackage{hyperref}
\usepackage{cleveref}
\usepackage{xcolor}
\usepackage{enumitem}
\usepackage{ulem}

\hypersetup{hidelinks}

\newcommand*{\icon}[1]{\scalerel*{\includegraphics{#1}}{\strut}}

\newcommand{\eg}{\emph{e.g.}}

\newcommand{\added}[1]{{\color{black}{#1}}}
\newcommand{\minor}[1]{{\color{black}{#1}}}
\newcommand{\tname}{\emph{SPROUT}}

\newcommand{\metaphor}{{\tname} is derived from{\itshape ``\textbf{S}tep-by-step \textbf{pro}gramming t\textbf{ut}orial authoring tool''}, with the metaphorical representation of the thoughts generating process of Large Language Models, symbolizing the system's goal to enable users a more controllable experience by utilizing the interactive visualizations of this process.}

\begin{document}

\graphicspath{{figs/}{figures/}{pictures/}{images/}{./}}

\title{SPROUT: an Interactive Authoring Tool for Generating Programming Tutorials with the Visualization of Large Language Models}

\author{Yihan Liu, Zhen Wen, Luoxuan Weng, Ollie Woodman, Yi Yang, Wei Chen.
\thanks{
      Yihan Liu, Zhen Wen, Luoxuan Weng, Ollie Woodman, Yi Yang, and Wei Chen are with the State Key Lab of CAD\&CG, Zhejiang University.
  	E-mail: \{lyh1024\,$|$\,wenzhen\,$|$\,lukeweng\,$|$\,orw\,$|$\,yang-yi\,$|$\,chenvis\}@zju.edu.cn.
  }
\thanks{  Wei Chen is the corresponding author.}
\thanks{Manuscript received April 19, 2021; revised August 16, 2021.}}

\markboth{Journal of \LaTeX\ Class Files,~Vol.~14, No.~8, August~2021}%
{Shell \MakeLowercase{\textit{et al.}}: A Sample Article Using IEEEtran.cls for IEEE Journals}

\IEEEpubid{0000--0000/00\$00.00~\copyright~2021 IEEE}

\maketitle

\begin{abstract}
The rapid development of large language models (LLMs), such as ChatGPT, has revolutionized the efficiency of creating programming tutorials.
\added{LLMs can be instructed with text prompts to generate comprehensive text descriptions for code snippets provided by users.}  
However, the lack of transparency in the end-to-end generation process has hindered the understanding of model behavior and limited user control over the generated results.
To tackle this challenge, we introduce a novel approach that breaks down the programming tutorial creation task into actionable steps.
By employing the tree-of-thought method, LLMs engage in an exploratory process to generate diverse and faithful programming tutorials.
We then present {\tname}, an authoring tool equipped with a series of interactive visualizations that empower users to have greater control and understanding of the programming tutorial creation process.
A formal user study demonstrated the effectiveness of {\tname}, showing that our tool assists users to actively participate in the programming tutorial creation process, leading to more reliable and customizable results.
By providing users with greater control and understanding, {\tname} enhances the user experience and improves the overall quality of programming tutorial. A free copy of this paper and all supplemental materials are available at \url{https://osf.io/uez2t/?view_only=5102e958802341daa414707646428f86}.
\end{abstract}


\input{chapters/1-intro}

\input{chapters/2-rw}

\input{chapters/3-formative}

\input{chapters/4-prompt}
\input{chapters/5-system}

\input{chapters/6-eval}

\input{chapters/7-user}
\input{chapters/8-fw}

\input{chapters/9-con}

\section*{Acknowledgments}
We would like to thank Jiehui Zhou and Minfeng Zhu for their
heartwarming support. We thank
anonymous reviewers for their insightful reviews. This paper is supported by the National Natural Science Foundation of China (62132017, 62302435), Zhejiang Provincial Natural Science Foundation of China (LD24F020011) and ``Pioneer'' and ``Leading Goose'' R\&D Program of Zhejiang (2024C01167).



\bibliographystyle{IEEEtran}
\bibliography{myref.bib}












\newpage



\begin{IEEEbiography}
[{\includegraphics[width=1in,height=1.25in,clip,keepaspectratio]{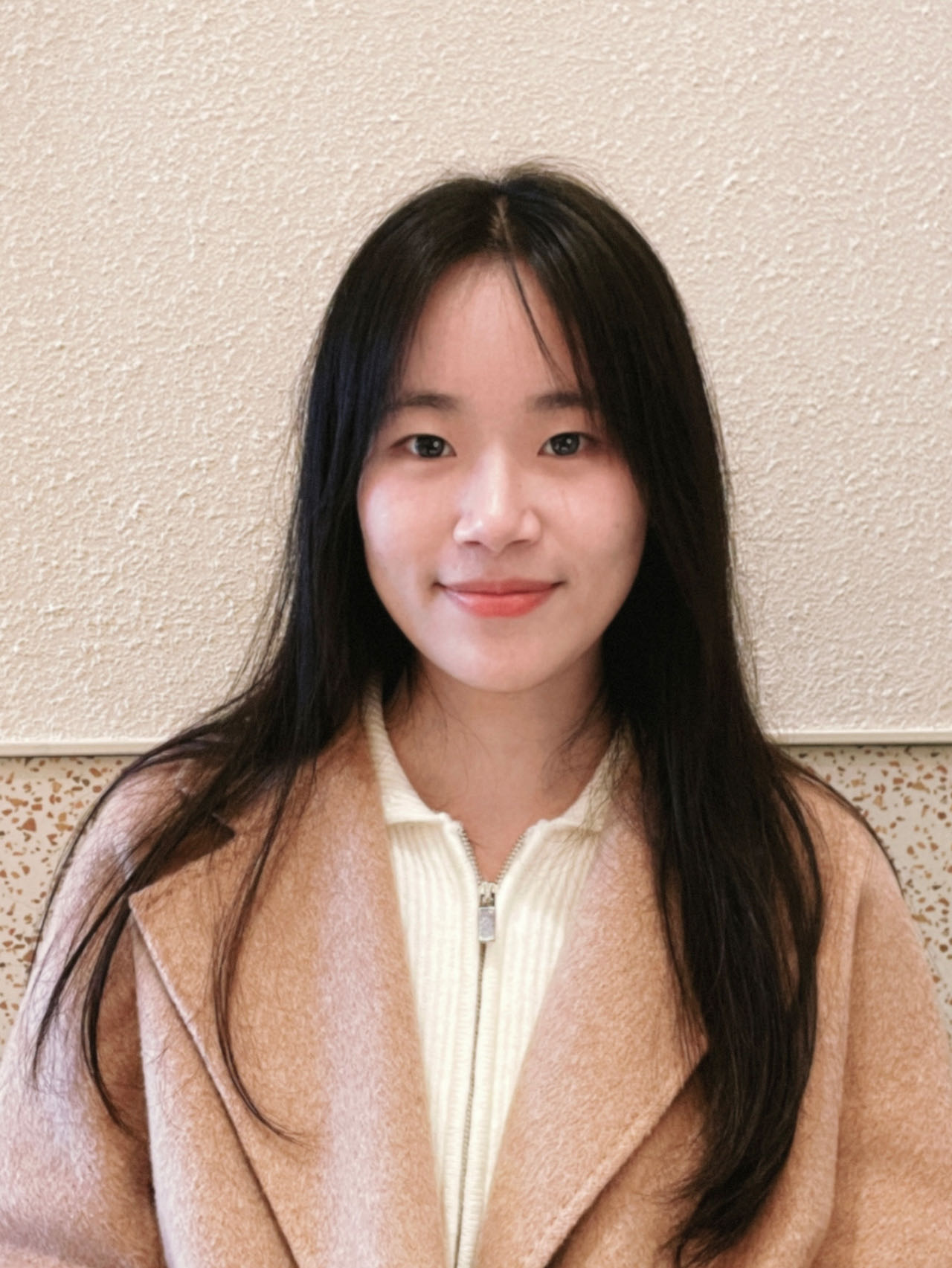}}]{Yihan Liu} received the B.E. degree in computer science and technology from the Zhejiang University, China, in 2022. She is currently  working towards the master's degree at the State Key Lab of CAD\&CG, Zhejiang University. Her research interests include visual analytics, human–computer interaction, and software engineering.
\end{IEEEbiography}

\begin{IEEEbiography}
[{\includegraphics[width=1in,height=1.25in,clip,keepaspectratio]{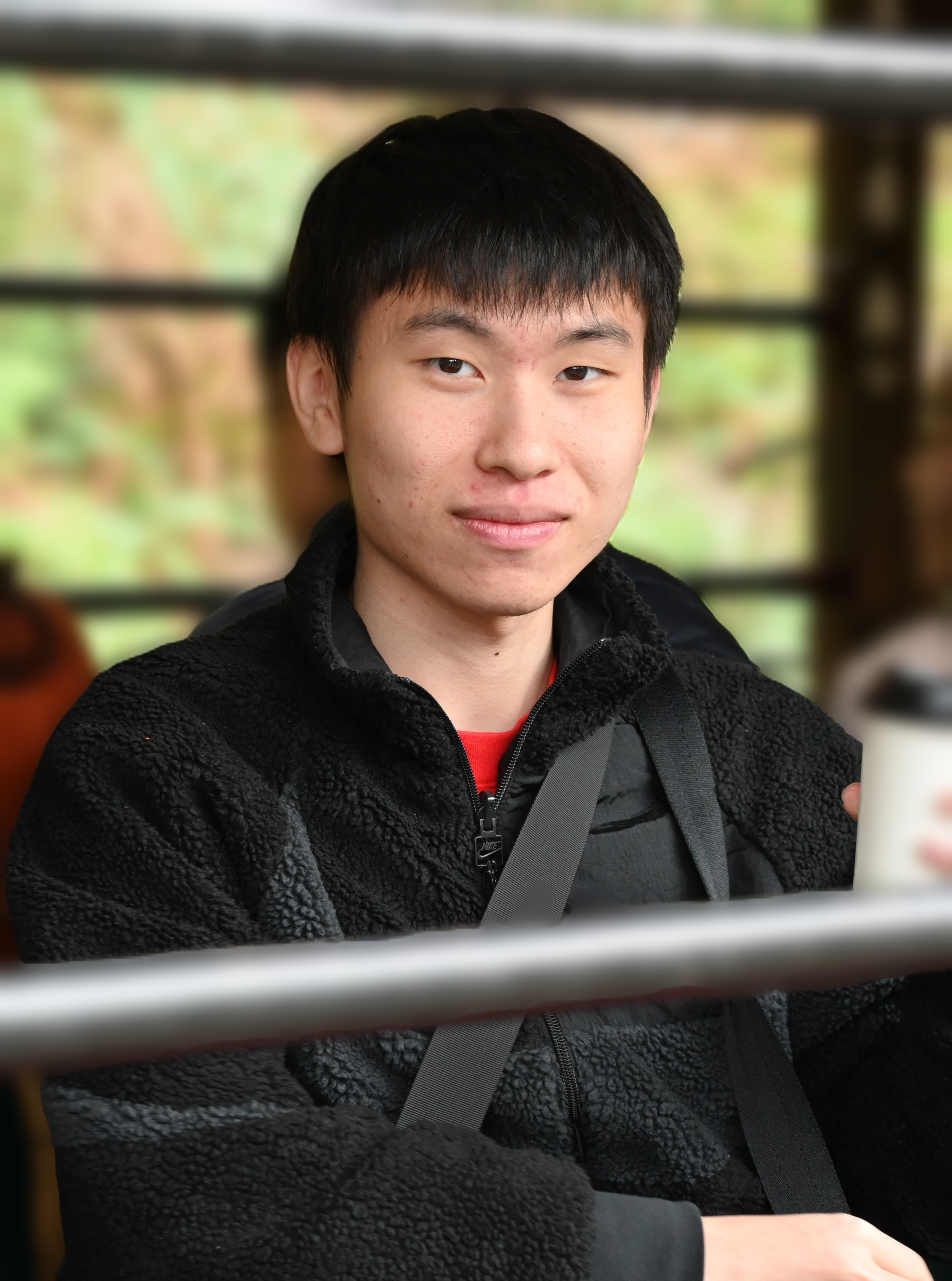}}]{Zhen Wen} is  a Ph.D. candidate at the State Key Lab of CAD \& CG, College of Computer Science and Technology, Zhejiang University. He received the B.E. degree in Software Engineering from Zhejiang University of Technology in 2021. His research interests include visual analytics, software engineering, and quantum computing.
\end{IEEEbiography}

 \begin{IEEEbiography}
[{\includegraphics[width=1in,height=1.25in,clip,keepaspectratio]{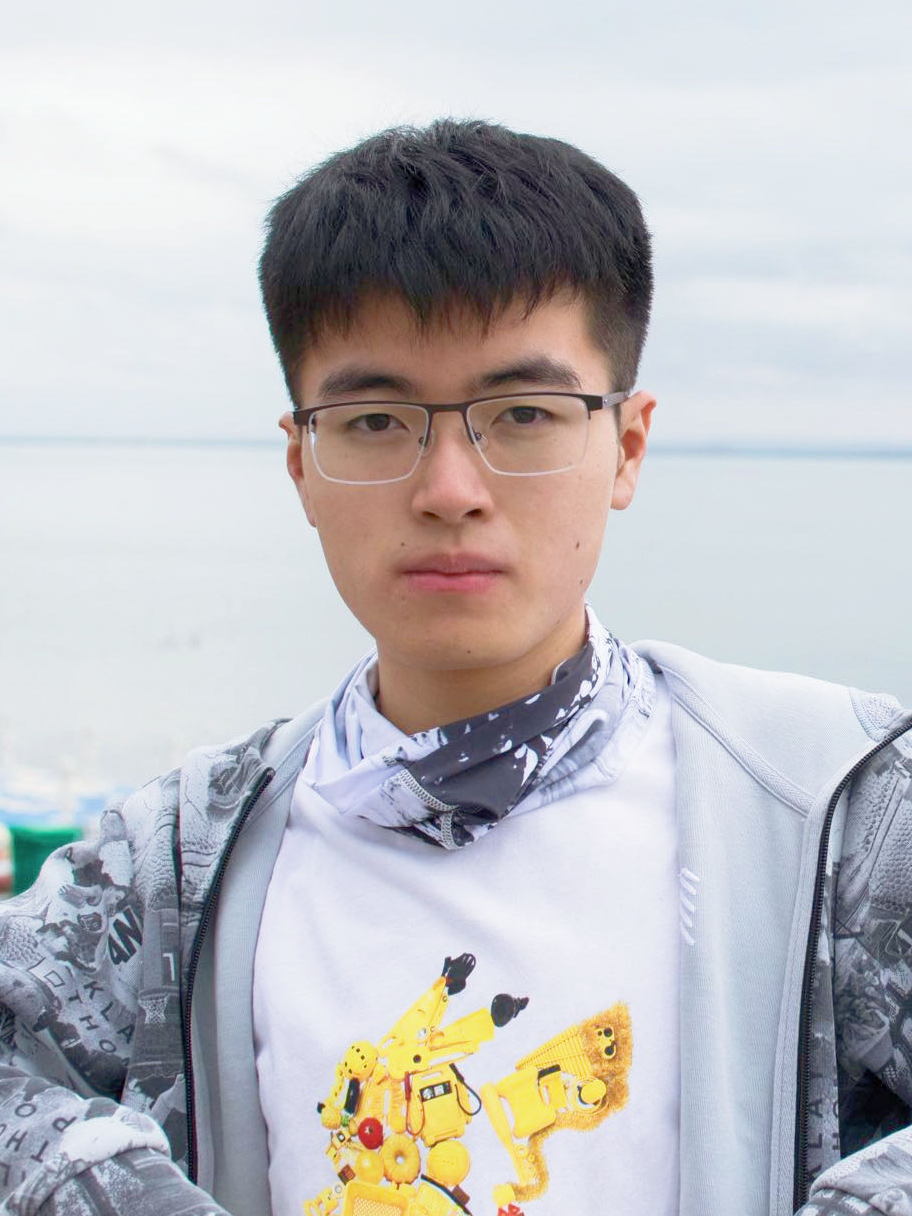}}]{Luoxuan Weng}
 is currently a Ph.D. candidate in the College of Computer Science and Technology, Zhejiang University, where he also received his B.E. degree in Software Engineering in 2021. His research interests include visual analytics, human-centered computing, and natural language processing.
\end{IEEEbiography}


\begin{IEEEbiography}
[{\includegraphics[width=1in,height=1.25in,clip,keepaspectratio]{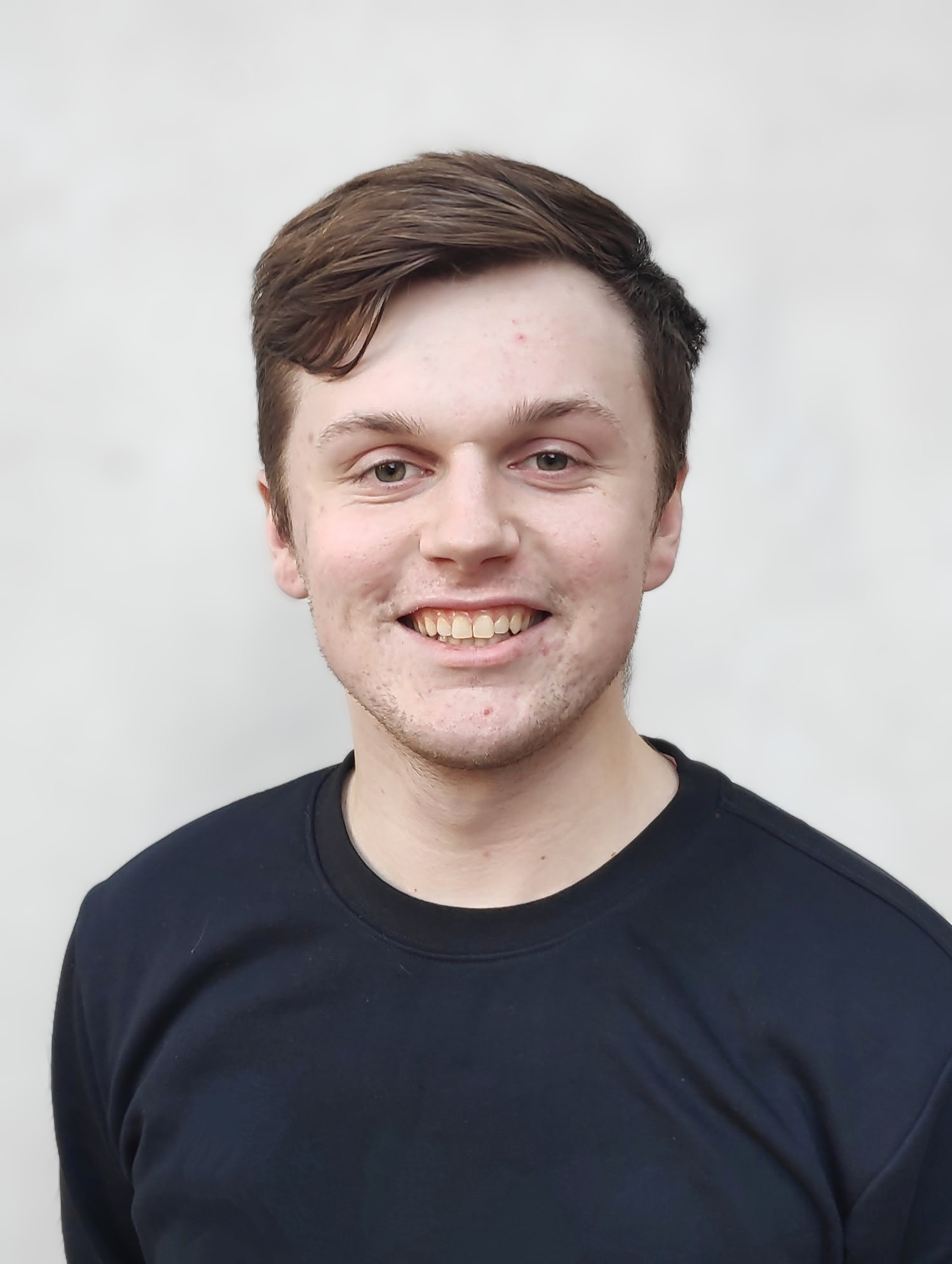}}]{Ollie Woodman}
is a Masters Student at the State Key Lab of CAD \& CG, College of Computer Science and Technology, Zhejiang University. He obtained a Bachelors of IT in Software Development and a Bachelors of Arts in Chinese Studies at Monash University in 2022. His research interests include large language models, software engineering and machine learning.
\end{IEEEbiography}


\begin{IEEEbiography}[{\includegraphics[width=1in,height=1.25in,clip,keepaspectratio]{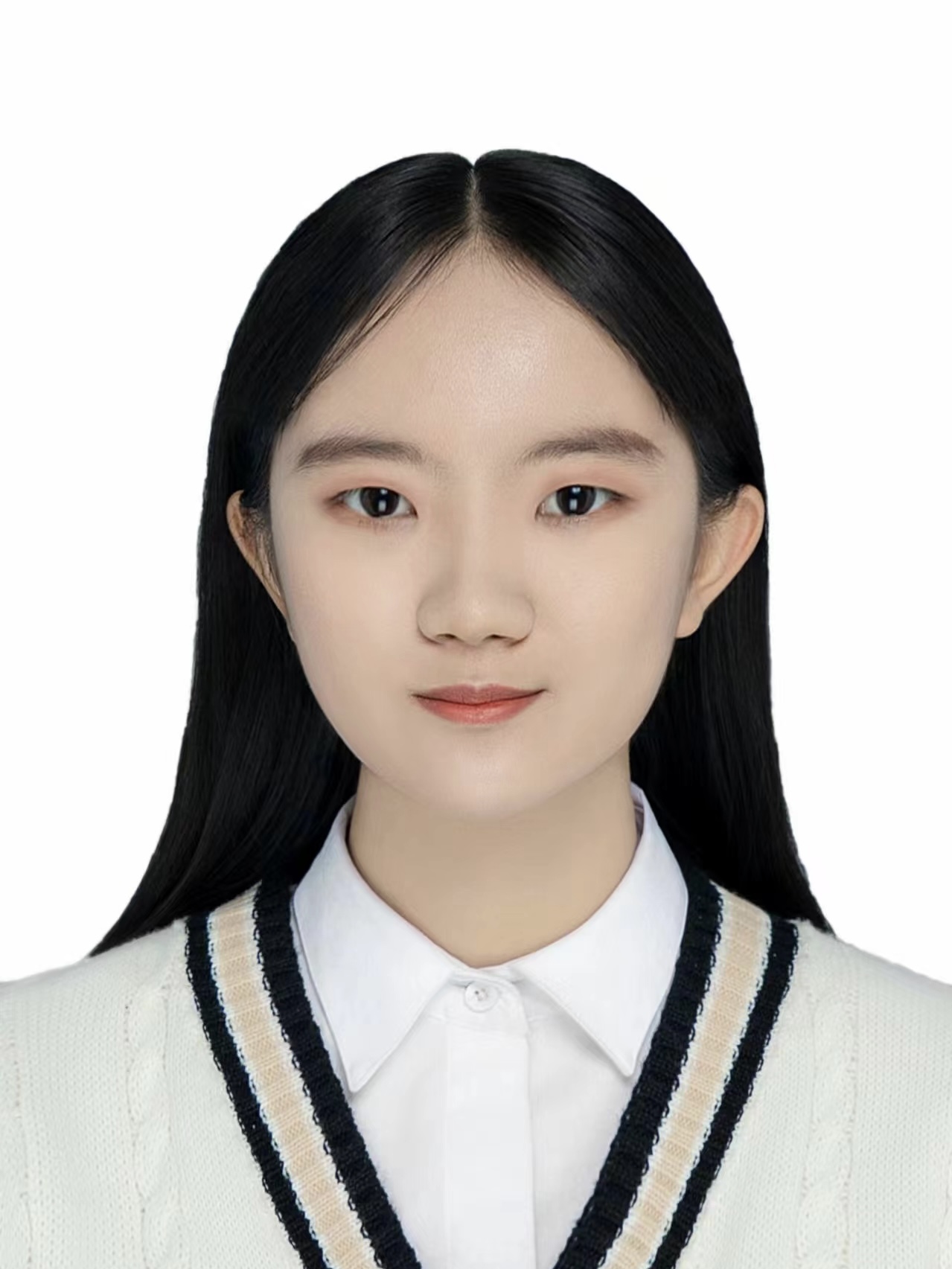}}]{Yi Yang}
is currently a Ph.D. candidate in the College of Computer Science and Technology at the Zhejiang University. She received the B.E. degree in Computer Science and Technology from the China University of Mining and Technology in 2023. Her research interests include computer vision and deep learning.
\end{IEEEbiography}

\begin{IEEEbiography}[{\includegraphics[width=1in,height=1.25in,clip,keepaspectratio]{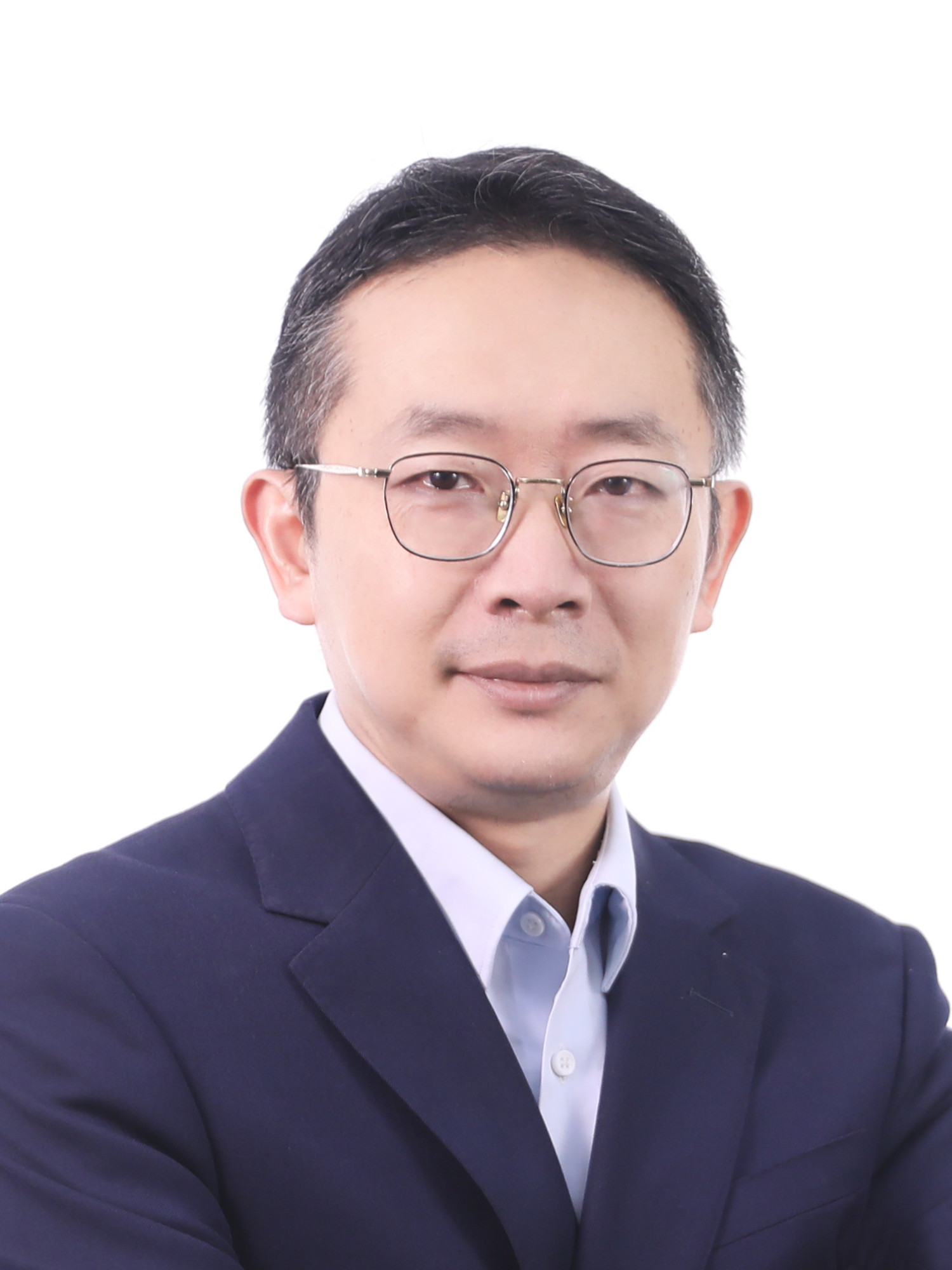}}]{Wei Chen}
is a professor in the State Key Lab of CAD\&CG at Zhejiang University. His current research interests include visualization and visual analytics. He has published more than 80 IEEE/ACM Transactions and IEEE VIS papers. He actively served in many leading conferences and journals, like IEEE PacificVIS steering committee, ChinaVIS steering committee, paper cochairs of IEEE VIS, IEEE PacificVIS, IEEE LDAV and ACM SIGGRAPH Asia VisSym. He is an associate editor of IEEE TVCG, IEEE TBG, ACM TIST, IEEE T-SMC-S, IEEE TIV, IEEE CG\&A, FCS, and JOV. More information can be found at: http://www.cad.zju.edu.cn/home/chenwei.
\end{IEEEbiography}



\vfill

\end{document}

%% file: chapters/1-intro.tex



\section{Introduction}

\IEEEPARstart{P}{rogramming} tutorials are used to teach learners coding skills and how to solve programming problems.
Numerous programmers disseminate their programming expertise through authoring various tutorials\cite{Pedagogical_2017_Kim}, such as blog articles, online videos, programming games, etc.
As the demand to learn programming has expanded, there has been a rise in the number of tutorial authors (\eg, developers\cite{FRAPT_Jiang_2017}, computer science educators\cite{RecurTutor_Hamouda_2019}, data scientists\cite{DocMatters_Wang_2022}, etc.), and the research on improving authoring experiences and the workflow of programming tutorials.


The recent prevalence of \textbf{L}arge \textbf{L}anguage \textbf{M}odels (LLMs) has revolutionized the conventional process of authoring articles \cite{gero_2022_sparks,sultanum_2023_datatales,biswas_2023_chatgpt}.
Similarly, programming tutorial authors can provide LLMs with code snippets and text instructions to generate tutorials.
The prominent ability of LLMs diminishes the effort for authors to author programming tutorials. 

However, the text generated by LLMs is not always perfect or error-free\cite{Ji_2023_Survey}.
Sometimes the exhaustive LLM-generated content is also not aligned with users' expectations.
These problems call for further refinement on the auto-generated content.
Owing to the inherent non-transparency of the generation process of LLMs, users lose access to the intermediate steps of the creation, leading to limited control and customizability when authoring articles~\cite{sultanum_2023_datatales}.
Additionally, to achieve ideal modifications often necessitates users' continuous interactions with LLMs via text-based conversation interfaces, which can not adequately support the complex requirements of users seeking to fine-tune tutorials.
It is even more time-consuming when dealing with programming tutorials, as it requires precise verification for the correctness and consistency.

The aforementioned limitations pose challenges in the process of authoring programming tutorials using LLMs.
These challenges have an impact on the overall quality and reliability of LLM-generated programming tutorials.
Some previous works delve in training LLMs to generate code-related content\cite{Chen_2021_Codex, Xu_2022_systematic} or proposing advanced prompting methods to promote the generation quality of LLMs\cite{Brown_2020_Language, Wei_2022_CoT,ahmed_2023_improving}.
However, to tackle with programming tutorial task, these approaches are not out-of-box solutions for generating faithful content accurately under human's interventions.
Several interactive systems have also been developed to enhance the efficiency of interactions with LLMs.
While proving useful, these systems primarily concentrate on specific needs, such as creative writing or data-driven article authoring\cite{yuan_2022_wordcraft, sultanum_2023_datatales}.
Unfortunately, these approaches are not suitable for the context of programming tutorial authoring, which necessitate accurate information organized in a more coherent way aligning with the source code.
Therefore, novel approaches are needed that account for the need to generate customizable programming tutorial and guarantee its faithfulness and consistency with the source code.






Recent studies propose instructing LLM to \emph{``think step by step''} to enhance the faithfulness and consistency of the generated results\cite{Yao_2023_ToT, Zhang2023VISARAH}.
Inspired by this, we decompose the tutorial authoring tasks into a step-by-step exploratory process towards final document accomplishment.
This approach has potential to enables users to attain a lucid comprehension and control over the intermediate steps throughout the entire creation process.
\added{
Thus, this study investigates methods to enhance user comprehension and control over \minor{the text generation process with LLMs}, with the ultimate goal of improving the quality and experience of authors.
For the sake of better understanding the underlying challenges in the authoring process,} 
we conducted a formative study with 6 experienced tutorial authors to have a grasp of the writing process and the challenges they encounter when transitioning from their conventional workflow to the authoring workflow with LLMs.
Participants reported that they can not effectively engage in the creation process using typical conversation interfaces \minor{due to the difficulty of controlling the generated content solely with text. The lack of visual cues between source code and generated text also led to }the overhead in comprehending and verifying the tutorial.

Consequent upon the the findings of formative study, we designed {\tname}\footnote[1]{\metaphor} ({\itshape \textbf{S}tep-by-step \textbf{pro}gramming t\textbf{ut}orial authoring tool}), an interactive system that breaks down the tutorial creation task into actionable steps. 
These steps are accompanied by interactive visualizations that empower users to have a greater control and understanding over the exploratory process, ending with more accurate and reliable results.
{\tname} adopts tree-like prompting strategies to allow LLMs to generate diverse and faithful tutorial content more reasonably, which serves as a entry point for subsequent user interventions to make accurate adjustment on tutorial in different aspects.
A series of purpose-specific interactions are also provided on the basis of the real-time visualizations of \minor{the text generation process with LLMs},  aiding users in crafting the tutorial from multiple levels.
To facilitate an intuitive understanding of the generated content, {\tname} leverages the inherent connections between the source code and tutorial, which are distilled by LLMs and demonstrated with visual representations to users, enabling the explicit comprehension of the creation process and model's behavior.
To evaluate the effectiveness of {\tname}, we conducted a technical evaluation and a user study.
The technical evaluation demonstrated that {\tname} achieved accurate text-code connections.
The user study revealed that {\tname} effectively assisted users in generating, modifying, and understanding programming tutorials with LLMs, ultimately leading to more reliable and customized results.
In summary, this paper makes the following contributions:
\begin{itemize}
    \item Understanding of the challenges inherent in the process of authoring programming tutorials with LLMs.
    \item {\tname}, an interactive system that utilizes novel prompting strategies and interactive visualizations to facilitate the step-by-step generation of programming tutorials with LLMs.
    \item A technical evaluation demonstrating the effectiveness of our prompting methods and a user evaluation confirming the usability of {\tname} and its support for the programming tutorial authoring workflow.
    
\end{itemize}

%% file: chapters/2-rw.tex
\section{Related Work}
\subsection{Programming Tutorial Authoring Tools}
Tutorials are a prevalent medium through which developers disseminate coding and programming knowledge.
Typically, a tutorial consists of source code, textual explanations, code examples, and multimedia components such as images and videos\cite{Composing_2020_Head}.
Crafting a high-quality tutorial necessitates consideration of various perspectives\cite{Pedagogical_2017_Kim, Typical_2014_Tiarks, QA_2012_Nasehi}. Even experienced authors require significant time to structure and present the tutorial content.
Therefore, numerous tools have been designed to improve the tutorial authoring process.
In-editor plugins like JTutor\cite{JTutor_2004_Kojouharov} and JTourBus\cite{JTourBus_2007_Oezbek} facilitate the construction of Java tutorials within code editors.
Some studies leverage interaction provenance collected from operating systems\cite{Torta_2017_Mysore} or web-based editors\cite{ITSS_2022_Ouh} to capture the programming process and produce tutorials. 
For example, VT-Revolution\cite{VT_2019_Bao} offers interactive video tutorial creation based on user actions, while Chat.codes\cite{Chat_2018_Oney} promotes collaborative code annotations and explanations.
Other works support users in drafting code examples by propagating changes to multiple code versions\cite{MultiStage_2013_Ginosar} or extracting concise code using a mixed-initiative strategy\cite{CodeScoop_2018_Head}.


Among the many forms of tutorials, the step-by-step structure is particularly accessible for novice authors due to its straightforward nature. Also, describing code as \emph{``a tour through the source code''}\cite{JTourBus_2007_Oezbek} proves beneficial for beginners in programming.
Therefore, our primary emphasis is on the authoring of textual content in step-by-step programming tutorials, where LLMs exhibit optimal performance.
We investigate the pain points in the authoring process and leverage LLMs to assist users in efficiently generating, modifying and understanding tutorial content, which furthers the line of research on programming tutorial authoring.

\subsection{LLMs for Document Generation}


The remarkable generative capabilities of LLMs have spurred the proliferation of LLM-based applications across various domains in writing scenarios.
In the domain of code-related content generation, several studies investigate the potential of LLMs in crafting code explanations by comparing LLM-generated content with that sourced from students\cite{leinonen_2023_comparing} or previous approaches\cite{khan_2022_automatic}, and validate its accuracy and comprehensibility.
Some endeavors have integrated these auto-generated code explanations into educational scenarios.
For instance, MacNeil et al. incorporate LLM-generated code explanations within a web software development e-book\cite{macneil_2023_experiences}.
Other efforts utilize LLMs to generate a myriad of CS learning materials including programming assignments and multifaceted code explanations\cite{macneil_2022_automatically,macneil_2022_generating}.
Additionally, Sarsa et al. explore LLMs' abilities in creating programming exercises and code explanations, with evaluations indicating that the generated content is novel, sensible and ready to use in certain cases \cite{sarsa_2022_automatic}.
\added{There are also developer tools for flexible code comment generation, for instance, GitHub Copilot\cite{github_2023_copilot} provides comprehensive text content pertaining to the code context.
Some researchers have developed LLM-based agents \cite{hong_2023_metagpt,qian_2023_communicative} responsible for documenting in software engineering, showcasing substantial potential in authoring code-related content.}

Nonetheless, previous works mostly focus on generating high-quality content, assessing the thoroughness and accuracy of code explanations.
For effective presentation to readers, such content often necessitates additional refinements, either through manual modifications or carefully-designed prompts.
NLP researchers are actively examining diverse prompt strategies and frameworks to optimize LLMs' performance\cite{Yao_2023_ReAct, Yao_2023_ToT} or mitigate \emph{hallucinations}\cite{dhuliawala2023chainofverification, sennrich2023mitigating}, seeking to harness their full potential. Our work builds on top of these endeavors by employing advanced prompt strategies to more effectively incorporate LLMs into the programming tutorial authoring workflow, ending with reliable and high-quality tutorials that meet users' expectations.


\subsection{Visual Interfaces for LLMs}

Despite the impressive capabilities of LLMs, traditional text-based conversational interfaces present challenges \added{when it comes to complex tasks that require the integration and manipulation of diverse data types\minor{\cite{Zhang2023VISARAH, sultanum_2023_datatales, yen_2023_coladder, feng_2023_promptmagician,wang_2024_lave}}, as they lack the capacity for direct multimodal data manipulation and the visualization necessary for complex, dynamic content creation and information synthesis\cite{jiang_2023_graphologue, suh_2023_structured}.}
Therefore, a variety of visual interfaces have been developed to enhance user interactions with LLMs.
For instance, researchers have designed interactive interfaces for visual programming prompt chains to lower the barrier for non-AI-experts\cite{wu_2022_promptchainer,wu_2022_ai,Arawjo_2024_forge}.
\minor{Some works aim to assist users generate better prompt with fow-shot examples\cite{Jiang_2022_prompt}, coordinated visualizations\cite{aditi_2023_aid} or direct manipulation actions\cite{masson_2024_direct} in graphical interfaces.}
In terms of diverse output representations, Graphologue\cite{jiang_2023_graphologue} translates LLM responses into graphical diagrams, while Sensecape\cite{suh_2023_sensecape} constructs a multi-level abstraction for information exploration and sensemaking. Both of them facilitate the information-seeking process.
Other works focus on designing interfaces for specific writing tasks, such as email writing\cite{goodman_2022_lampost}, story creation\cite{yuan_2022_wordcraft}, argumentative writing\cite{Zhang2023VISARAH}, scientific writing\cite{gero_2022_sparks} and journalistic angle ideation\cite{petridis_2023_anglekindling}. These interfaces ensure users achieve ideal results through iterative interactions.
Notably, DataTales\cite{sultanum_2023_datatales} introduces more flexible and intuitive interactions for authoring data-driven narratives, while TaleBrush\cite{chung_2022_talebrush} utilizes line sketching interactions with GPT-based models to control and understand a protagonist’s fortune in co-created stories.
More recently, Kim et al. develop a framework for object-oriented interaction with LLMs, which can generalize to a wider range of writing tasks\cite{kim_2023_cells}.

Motivated by these prior studies, our work aims to facilitate the process of authoring programming tutorials with LLMs.
\minor{Despite the similarities to various writing tasks, the unique aspects such as writing style, the process of creation, the nature of data involved, and the necessity for consistency and faithfulness make their methodology unsuitable for the task of authoring programming tutorials.}
Through a set of interactive visualizations, our approach empowers users to actively participate in the generation and refinement of tutorials while maintaining flexibility and precision in their control on the process.
Additionally, we extract and present visual representations of the connections between the source code and the corresponding LLM-generated results, facilitating users’ explicit understanding of the tutorial content.

%% file: chapters/3-formative.tex
\section{Formative Study}
\label{sec:formative-study}
We conducted a formative interview study to analyze the challenges encountered in tutorial creation and workflow with LLMs, from which we distilled four essential design requirements to improve the authoring process.

\subsection{Participants and Procedure}
\added{In order to comprehend the challenges and difficulties encountered across various tutorial authoring scenarios, six experienced tutorial authors were interviewed (age from 23 to 28).
Their wide-ranging expertise in authoring diverse tutorials for various users spans from novice to expert levels. 
Thus, the insights derived from their collective experiences provide a more comprehensive and universally applicable understanding, beneficial to programming tutorial authors of various proficiency levels.}
Four of them are experienced programmers (E1-4) and two are educators.
One of the two educators teaches computer science (E5), while the other instructs competitive programming (E6).
All of them have more than three years of experience authoring programming tutorials.
In addition, they all have used ChatGPT in recent months.
Before the interview, we collected twenty text-based programming tutorials from various sources, including Stack Overflow, Wikipedia, GeeksforGeeks, etc., as well as those auto-generated by LLMs.
In our formal interviews, we first inquired about the authors' writing process and commonly-used tools for creating tutorials, then presented them with all of the collected tutorials and solicited their opinions on the advantages and downsides of each.
Afterwards, participants were asked to create programming tutorials utilizing ChatGPT and document editors.
Finally, we collected feedback on their authoring experience, focusing on how they utilized LLMs in their workflow and what made it challenging to create high-quality tutorials in this process.
The interviews were conducted through online meetings and lasted 45 to 60 minutes.

\subsection{Findings}
All participants tried to author programming tutorials with LLMs. They provided the source code, retrieved the resulting tutorial, and iteratively modified it by seeking ChatGPT's assistance.  
We found that they usually managed to generate different versions with multiple categories of content to enrich the tutorial.
\minor{The five most common categories among them are title, background, code explanation, notification, and summary.}
\minor{The behavior analysis indicated a frequent tendency among participants to leverage LLMs for generating content, structuring the tutorial, and polishing details.}
However, during the tutorial authoring process with LLMs, participants commonly encountered challenges that impacted production efficiency.

\added{
\textbf{Difficult to generate tutorials that accurately reflect diverse coding scenarios.}
This challenge stems from a need to encompass a wide array of programming paradigms, languages, and problem-solving techniques within instructional content.
It potentially creates gaps in understanding for learners who come from different experience levels or who are looking to apply their skills to a variety of contexts.
As such, during the authoring process with ChatGPT, participants often complained about the difficulty to obtain tutorials aligning with their mental model.
For example, E5 hoped that ChatGPT could only explain the crucial parts of the code since the tutorial is for experienced programmers, but only found the output elaborated on some unimportant details, calling for additional effort in modifying the prompt to generate tutorials again.
Some participants proposed the desire to obtain tutorials with different frameworks, which can \textit{``provide more inspiration when having no ideas about what to do next''} (E4).
Apart from that, during the iterative tutorial authoring process, some participants noted that the content generated by ChatGPT occasionally deviated from being faithful to the original source code. 
It is significant to keep the tutorial consistent with the original source code.
Such experiences underscore the need for reasonable prompting strategies for generating rich and reliable content in diverse writing scenarios with different requirements.
}

\added{\textbf{Lack of flexible interactions for tutorial generation and modification.}}
As auto-generated content seldom fully met participants' expectations, they had to iteratively modify the tutorial content during the exploration.
Some of them sought to simplify this process using ChatGPT, but only to find it \textit{``inconvenient and confusing''} (E6).
E6 reported that combining two paragraphs to shorten the document necessitated a meticulously crafted prompt, otherwise ChatGPT's response was merely \textit{``a non-sensible mess''} (E6).
Additionally, participants mentioned concerns over the tediousness of constantly \textit{``switching back and forth between the editor and ChatGPT''} (E4), crafting prompts, and ensuring that the new content seamlessly integrated into the existing context.


\textbf{Significant overhead for understanding and verifying the tutorial.}
The linear conversation interface of ChatGPT, where user inputs and LLM outputs appear sequentially, becomes problematic due to the lack of visual linkages between code and text, particularly in multi-step interactions with iterative refinement.
They usually needed to \textit{``browse the whole document''} (E5) to discern which code segment each paragraph is describing, as ChatGPT omitted code references.
The situation was exacerbated during modifications, as responses often excluded the original content and code, forcing participants to \textit{``painstakingly match code snippets to their text because of their separate presentations''} (E3).
Moreover, the independently showcased nested code often lost its proper indentation, pushing participants to trace back to its context. This caused \textit{``additional mental overload to recognize the correct execution scope''} (E2).
Collectively, participants felt that these unpredictabilities \textit{``somewhat negated the intended efficiency gains from Large Language Models''} (E6).

\subsection{Design Requirements}
To tackle the problems concluded above, we aim to implement an interactive system to improve the authoring process with LLMs and obtain desirable programming tutorials.
The \textbf{D}esign \textbf{R}equirements can be summarized as follows:

\textbf{DR1. Generating diverse and faithful content.}
In programming tutorial authoring, users tend to utilize LLMs to generate a wide range of content, including variations in structure, level of detail, and other aspects, in order to inspire their writing ideas.
Simultaneously, they require the generated content aligns precisely with the intended functionality and logic.
Therefore, the system should generate diverse content while maintaining its faithfulness to the original source code provided by users.



\textbf{DR2. Supporting precise and in-context modification.}
Users often engage in iterative interactions with LLMs through text-based conversations to refine the generated tutorial content according to their preferences, \minor{mainly focusing on the aspects of tutorial structure and content details.}
However, even slight modifications in the input prompts can result in significant and unforeseen changes in the entire generated output.
To maintain the coherence and accuracy of the final output, the system should support users in making precise modifications while preserving the contextual integrity of the existing generated tutorial content.

\textbf{DR3. Maintaining explicit connection between inputs and outputs.}
Users usually need to carefully read the source code and outputs with the purpose of understanding and verifying the LLM-generated content.
However, in linear conversation interfaces, this task becomes time-consuming and challenging.
Additionally, the lack of transparency and traceability between user inputs and model outputs can result in inadequate trust and confidence towards the final tutorial.
Therefore, in this system, the tutorial content should be augmented with explicit visual representations of connections between the generated content and its related code snippets, which should be extracted and maintained by the system.
This allows users to easily comprehend and verify the tutorial and the reasoning behind it.

\textbf{DR4. Providing intuitive and flexible interactions.}
Utilizing natural language as the channel to communicate with LLMs poses challenges for users to clearly convey their requirements and ideal output.
In order to acquire qualified and satisfactory tutorials, users have to craft them continually, by some cumbersome editing operations and iterations on prompts for generation and modifications on the draft from ChatGPT.
Therefore, to circumvent these issues, the system should be synthesized with intuitive and flexible interactions to eliminate users' efforts for manual adjustment on content or revision on prompts,
\added{for instance, enable direct manipulation of the source code or generated textual content.
This allows authors to focus more intently on the programming tutorial itself.}


%% file: chapters/4-prompt.tex
\begin{figure*}[tb]{
    \centering
    \includegraphics[width=1\linewidth]{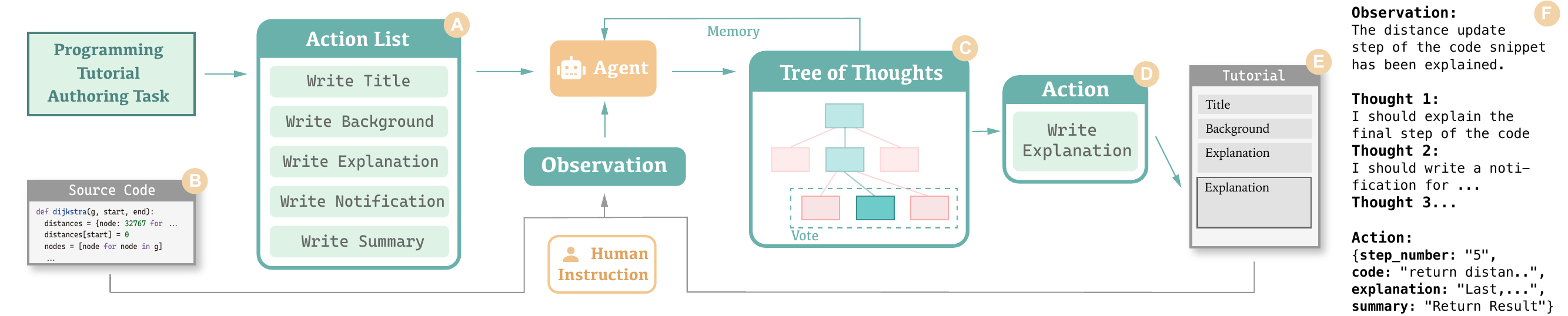}
    \caption{%
      The framework of {\tname}'s prompt strategies.
      We break down the programming tutorial creation task into actionable steps (A) which are provided as initial system prompts to the agent.
      During the generation process, the agent receives the source code (B) and user instructions, generating multiple potential thoughts (C) based on its memory and the observation derived from the source code and current tutorial content (E).
      Then it takes the proper action (D) and plans for subsequent steps.
      The iterative process continues until the tutorial is complete.
      (F) shows an example iteration.
    }
    \label{fig:framework}
  }
  \end{figure*}
\section{Prompting for Programming Tutorial Authoring}
\label{sec:prompt}
A primary goal of our research is to make the LLM generation process more transparent and controllable.
To achieve this, we adopt the tree-of-thought (ToT) methodology\cite{Yao_2023_ToT} to prompt models.
It enables models to take intermediate steps towards completing the tutorial.
By adopting the ToT methodology, we allow models to explore multiple choices during generation process.
This promotes a more interactive and iterative approach to tutorial authoring, empowering users to have greater control over the generated content.
In this section, we present our proposed prompt design that combines the ToT methodology with the specific requirements of programming tutorial authoring.
\Cref{fig:framework} illustrates the framework of our prompt strategies.
Our prompt strategies are developed and tested on \verb|gpt-3.5-turbo| version of ChatGPT\cite{OpenAI_2023_ChatGPT}, one of the most popular LLMs.
Please refer to supplemental materials for original prompts.

\subsection{Generation of Thoughts}
\label{sec:prompt:generate}
We envision the LLM as an agent that takes on the role of making decisions on how to plan and execute steps to accomplish programming tutorials.
The agent is instructed to think in a step-by-step manner to generate faithful content, and expand thoughts in a tree-like structure to produce diverse tutorial content (DR1).

\textbf{Dividing task into actions.}
We break down the programming tutorial creation task into discrete and actionable steps.
This allows the agent to consider these steps as fundamental units of a plan.
Through the formative study (\cref{sec:formative-study}), we have identified 5 key components of tutorial content, including the title, background, code explanation, notification, and summary.
Consequently, we define a list of actions that correspond to writing different content within the tutorial, e.g., \verb|`write title'| or \verb|`write background'|.
The action list and the descriptions of actions are provided to the agent before solving tasks.

\textbf{Thinking step by step.}
To guide the decision-making process leading to faithful results, we instruct the agent to think step-by-step following the ReAct (Reason and Act) paradigm\cite{Yao_2023_ReAct}, which requires the agent to solve the task in a specific order: \textit{observation}, \textit{thought}, \textit{action}.
At each step, the agent first presents the observation derived from the source code and the current content of tutorial.
Subsequently, it generates the thought of the appropriate next action to be taken with a reasonable justification.
The agent then takes the action and continues to plan for the next step.
This process continues iteratively until the tutorial creation process is completed. 
This systematic approach allows the agent to make informed decisions at each stage of the tutorial creation process, resulting in faithful and reliable outcomes.

\textbf{Expanding tree of thoughts.}
We integrate the tree-of-thought methodology in the decision-making process to enable the agent to generate multiple potential thoughts for consideration, leading to diverse content in the generated tutorial. 
For example, after explaining a specific code snippets, the agent may generate thoughts such as writing a notification highlighting common mistakes related to the code or directly providing a summary to conclude the tutorial.
Consequently, the agent votes on the generated thoughts and make a decision on the next action to take. 
To provide an exploratory space for authors, we maintain the tree of thoughts as a memory space for the agent.
Users are allowed to navigate through the tree of thoughts to explore multiple potential results of generation.

\subsection{Manipulation of Model Actions}
The above prompt strategies enables a completely automatic process to create tutorial without human intervention.
We propose to perform in-context interventions by inserting instructions in the gap of each step to manipulate the model actions.
This capability enables authors to make contextual adjustments to the tutorial materials, resulting in more tailored and customized programming tutorials (DR2).

In particular, we design prompt strategies to support authors in the \minor{process of generating content (generation), adjusting the structure of tutorial (organization), and refining the detail of content (refinement)}, which are derived from the formative study (\cref{sec:formative-study}).
\added{These strategies serve as underlying methods within features of our system (see \cref{sec:system})}:
\vspace{-2pt}
\begin{itemize}[leftmargin=10pt, noitemsep]
    \item \textbf{Generation:}
    ``\emph{The next step should be to write for $\langle$ code $\rangle$}''.
    This prompt is used to enable users to define a specific range of code to explain.
    It instructs the agent to thought closely associated with the specific code fragments when generating new content for tutorial.
    \item \textbf{Organization:}
    ``\emph{Explain $\langle$ code $\rangle$ in the next multiple steps}'' or ``\emph{Explain $\langle$ code $\rangle$ in one paragraph}''.
    These prompts are used to adjust the structure of the tutorial.
    They assist users in organizing the tutorial by separating the explanation of important code into multiple steps or summarizing less important code in a concise manner.
    \item \textbf{Refinement:}
    ``\emph{Explain $\langle$ code $\rangle$ in the style of $\langle$ style $\rangle$}'',
    ``\emph{Refine the last step $\langle$ `shorter' | `longer' $\rangle$}''.
    These prompts are used to refine the writing style or level of detail in a paragraph,
    which allows users to modify the tutorial content in a more granular manner.
\end{itemize}
\vspace{-2pt}
These prompt strategies empower authors to iteratively refine and customize the programming tutorial content according to their individual preferences and needs.

\subsection{Extraction of Text-Code Connections}
The relationship between the LLM-generated tutorial and its underlying code are not always clarified.
The extraction of text-code connections strategy focuses on identifying and establishing links between relevant textual explanations and corresponding code snippets (DR3). 

\begin{figure}[htbp]
  \centering
  \includegraphics[width=\columnwidth]{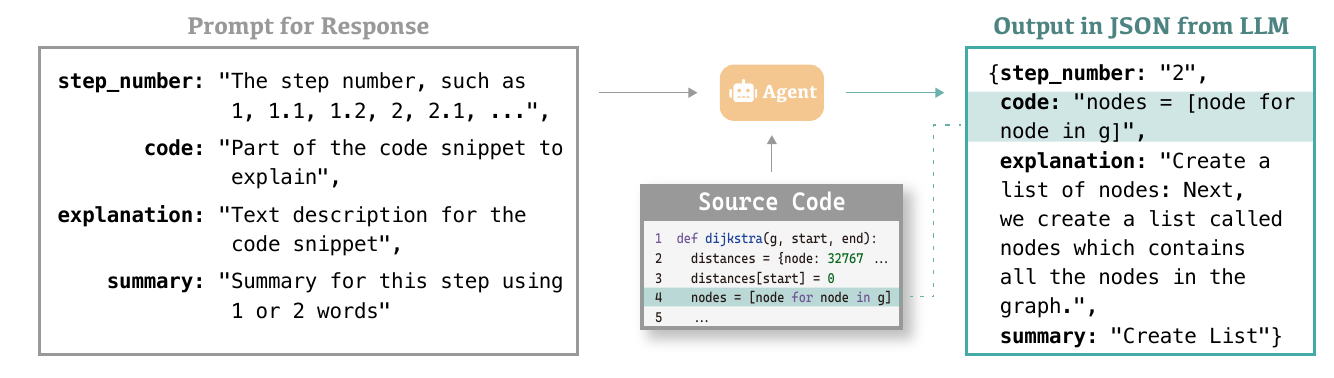}
  \caption{%
   An example output with text-code connection. The agent is instructed to provide text-code connection information within its response.
  }
  \label{fig:method-connect}
\end{figure}

We instruct the agent to present each code-related text accompanied with its connection to the code.
The agent is required to provide extra information when writing textual descriptions for a specific piece of code, e.g., \textit{step number}, \textit{code}, \textit{explanation}, and \textit{summary}.
We then link the reference code with the source code using string matching.
\added{\Cref{fig:method-connect} showcases an example of our method.}
This strategy is seamlessly integrated into the generation process, and also ensures that textual explanations are closely associated with the relevant code, enhancing the overall comprehensibility and effectiveness of the tutorial materials.
The tree of thoughts annotated with extracted connections helps authors structure their thoughts and logically sequence the tutorial materials.
It assists authors in creating and organizing tutorial content effectively, aiding authors in maintaining coherence and clarity throughout the tutorial.

%% file: chapters/5-system.tex
\begin{figure*}[tb]{
  \centering
  \includegraphics[width=\linewidth]{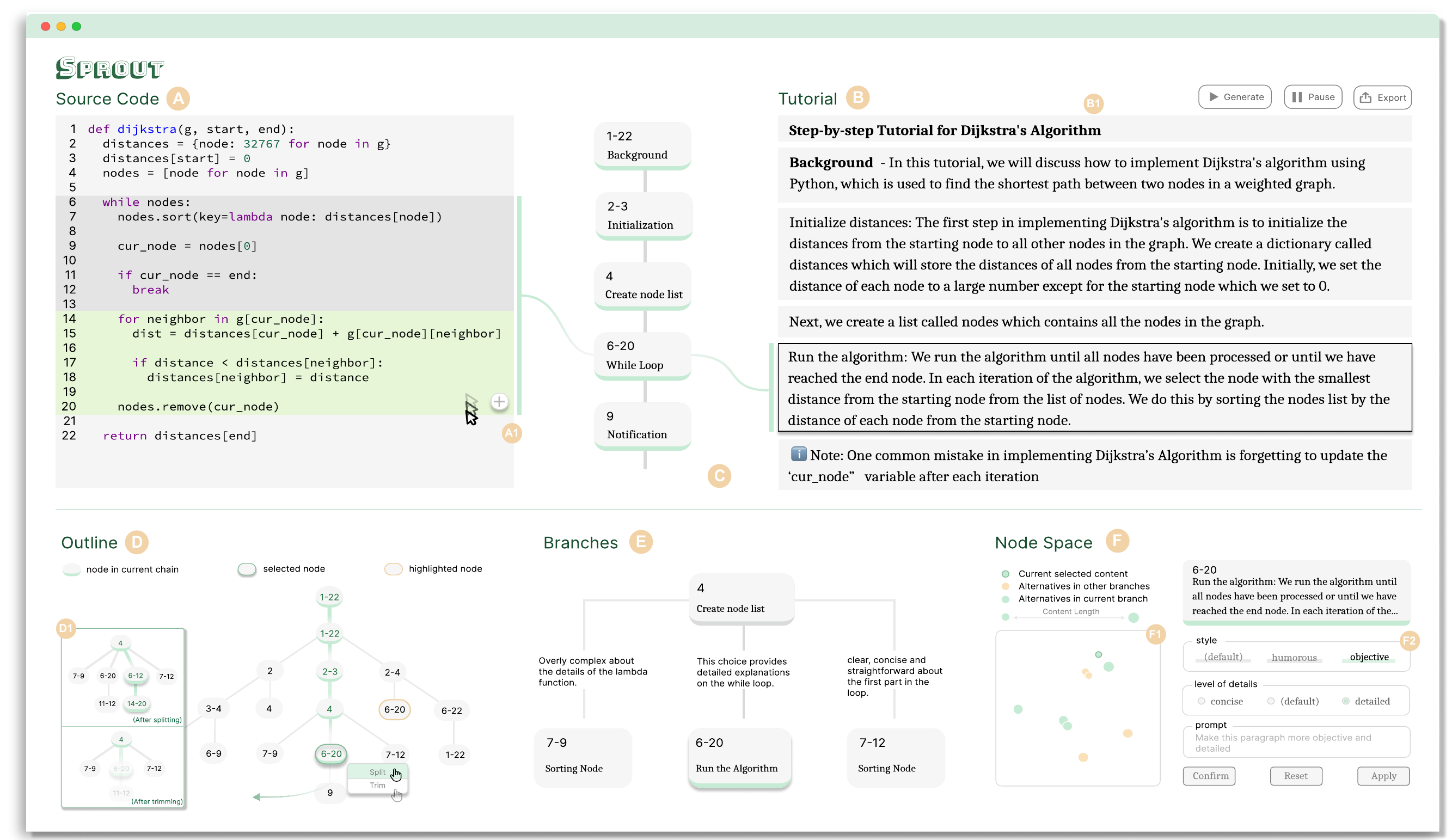}
  \caption{%
    The system interface of {\tname}.
    The Code View (A) displays the source code. 
    The Tutorial View (B) presents the LLM-generated tutorial content.
    Between them is the Chain View (C), showing the node chain of current paragraphs selected by users.
    The Outline (D), Branches (E), and Node Space (F) Views provide interactive visualizations for multi-level modifications such as elaborating, adjusting structure, and polishing content.
  }
  \label{fig:system}
}
\end{figure*}

\section{SPROUT}
\label{sec:system}
We develop \emph{SPROUT}, a prototype system that supports efficient and flexible step-by-step programming tutorial authoring with LLMs according to interactive visualizations of the intermediate generation process.
In this section, we first present an overview of the system, introducing the interface design and visual representations in \cref{sec:system:interface}. Next, we introduce a series of features that are designed for enhancing user experience in the process of tutorial authoring (\cref{sec:system:generate}-\cref{sec:system:vis}).



\subsection{Overview}
\label{sec:system:interface}
Our system consists of 6 views as demonstrated in {\cref{fig:system}}.
The \textbf{code view} (\cref{fig:system} A) allows users to enter the source code they intend to describe about.
The \textbf{tutorial view} (\cref{fig:system} B) displays the text content generated by LLM with a block-based structure, where a block presents a paragraph. 
Between the code view and tutorial view lies the \textbf{chain view} (\cref{fig:system} C), which serves as a connector, assisting users to navigate between code and tutorial.
Within this view, each node \icon{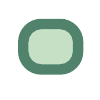} is a simplified visual representation of the discrete output units generated during the step-by-step creation process. Each node displays information about a specific paragraph and the corresponding code it describes.
The chain \icon{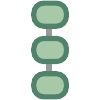} is concatenated by these nodes, which collectively represent the entire tutorial.

The \textbf{outline view} (\cref{fig:system} D) visualizes the intermediate generation steps of LLM in the form of a tree graph \icon{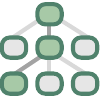}.
It consists of nodes~\icon{./icon/node} and branches \icon{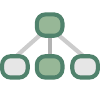}, illustrating the hierarchical structure of the LLM’s thought process.
The \textbf{branch view} (\cref{fig:system} E) enables users to switch between different branches \icon{./icon/branch} to explore alternative thought paths and make adjustments to the structure of generated tutorial. 
The \textbf{node space view} (\cref{fig:system} F) offers  alternative content options for a selected node. Users have the flexibility to choose from these alternatives and customize them in multiple dimensions.

\subsection{Content Generation from Code Snippets}
\label{sec:system:generate}
{\tname} supports two interactions that enable the generation of diverse and user-customized content from the source code provided by users (DR1, DR4).

\textbf{Agent generation.}
Sometimes users have limited knowledge about the code they want to describe or feel hesitated about the next steps to take.
In this situation, they can make the LLM-based agent shoulder the responsibility to sketch the framework.
In the tutorial view (\cref{fig:system}~B), users can click on the \textit{Generate} option to start the generation. 
{\tname} then automatically generates the tutorial content step by step, following the tree-of-thought methodology.
To ensure users' awareness and control over the tutorial generation process, {\tname} updates the tutorial content and renders the tree graph in real-time. 
Meanwhile, users can use the \textit{Pause} option to halt the generation process and review the current tutorial content, ensuring it aligns with their expectations, especially if the auto-generated content significantly deviates from what users anticipated.

\textbf{User-defined generation.}
In some cases, users have a clear intention regarding the code snippet they want to write about next.
However, they may find it challenging to effectively communicate their ideas in written form to inform the LLMs.
{\tname} saves users' arduous work to write and test various prompts to get a satisfactory response.
Instead, they can simply brush their target code snippet and click the floating button to effortlessly add new content to the current tutorial (\cref{fig:system}~A1). 
Supported by prompting strategies we proposed (\cref{sec:prompt}), {\tname} generates paragraphs related to the code with consideration of the coherence to the surrounding contexts.
And the node corresponding to the newly generated content is appended to the end of the current chain.
This ensures that the generated content is seamlessly integrated and follows a logical progression within the existing tutorial. 

\subsection{Tutorial Modification through Tree Graph}
{\tname} enables users to utilize the tree graph as the primary controller for flexible and intuitive modification of the tutorial content through manipulating the nodes within the tree (DR2, DR4).
This feature aims to alleviate the need for users to directly prompt the LLMs and reduces their burden.
For instance, users can consolidate multiple paragraphs to reduce their quantity, emphasize important code snippets by elaborating on them, and conveniently remove unnecessary parts from the tree graph for easier content management.

\begin{figure}[htbp]
    \centering
    \includegraphics[width=\columnwidth]{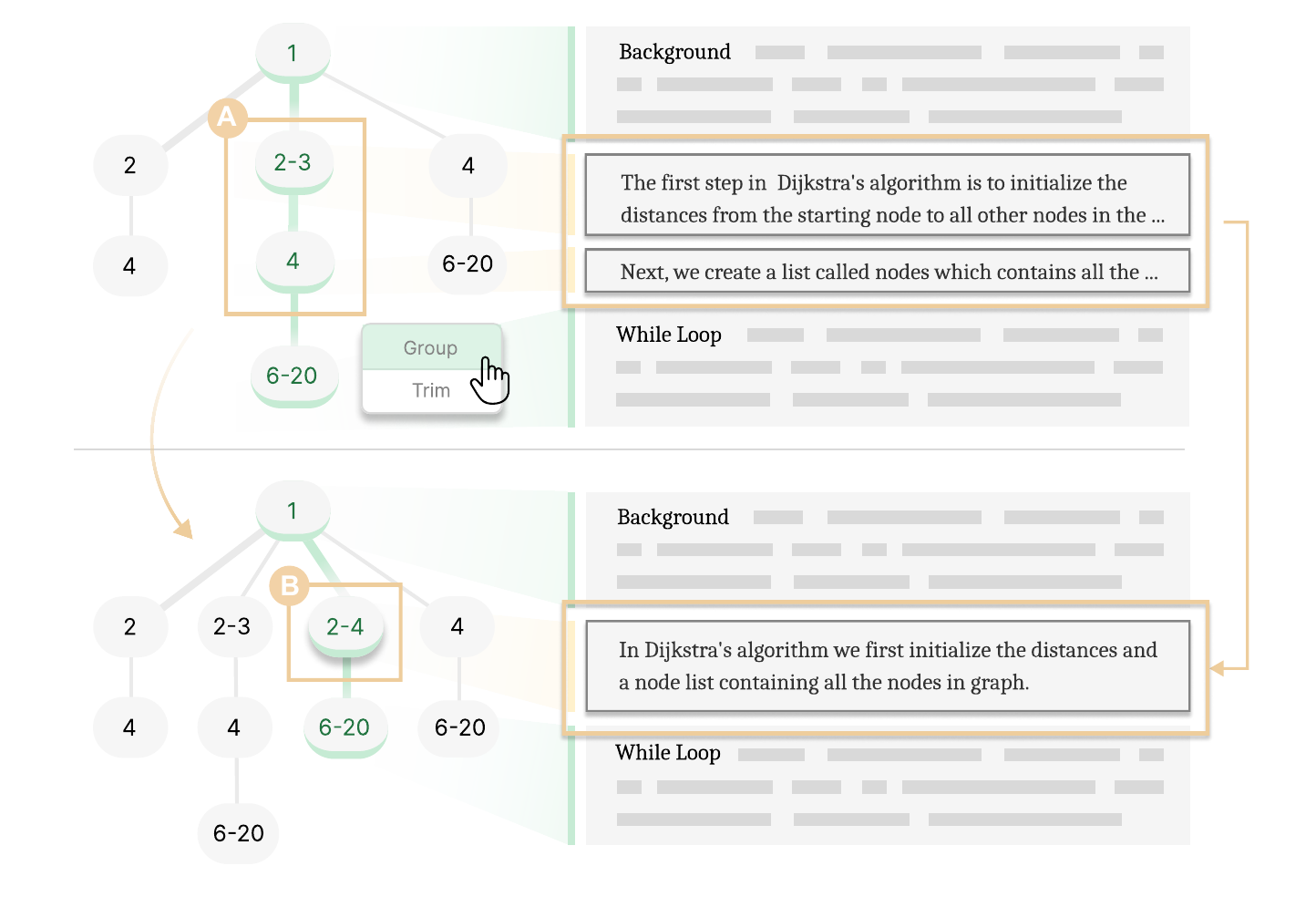}
    \caption{%
    Select two nodes (A) and group them into one (B).%
    }
    \label{fig:int-group}
\end{figure}

\textbf{Condensing through node grouping.}
Sometimes LLMs provide overly exhaustive content for users, resulting in unclear key points in the tutorial and overwhelming readers.
Users have to manually identify paragraphs that are less important or too familiar to the majority of readers and condense the content together in the typical workflow.
{\tname} thus supports node grouping in the outline view to simplify this process.
For example, when a user finds out that the content of two paragraphs can be integrated into one, they can select them in the tree graph and navigate to the \textit{Group} option to make the LLM achieve a seamless content combination (\cref{fig:int-group}).
The selected nodes will all be replaced by the newly-integrated node in a new branch, ensuring access to their old version.

\textbf{Elaborating through node splitting.} 
In addition to condensing, elaboration is another essential task for maintaining a balanced quantity throughout the entire tutorial.
When users explore the tree graph and discover that a key concept is mentioned but lacks sufficient explanation, they can select the corresponding node and navigate to the \textit{Split} option.
{\tname} prompts the LLM to generate additional content, which is then presented in new paragraphs.
Simultaneously, {\tname} creates a copy of the branch where the selected nodes are located, and the selected nodes are split into newly-generated nodes (\cref{fig:system}~D1).

\textbf{Deleting through node trimming.}
Since {\tname} showcases complete thinking paths of the agent when authoring the tutorial, the tree graph may accumulate redundant nodes as users iterate on the tutorial, making it difficult to navigate and manage effectively.
Therefore, {\tname} offers a node trimming feature that allows users to delete specific nodes in the tree graph that contain the undesired content.
When users pinpoint some redundant or unnecessary paragraphs, they can use the \textit{Trim} option to delete the corresponding nodes, and their children nodes will be removed from the tree at the same time (\cref{fig:system} D1).


\subsection{Context Switch across Branches}
\label{sec:system:branches}
To facilitate a more diverse (DR1) and flexible (DR4) organization of the tutorial's content, {\tname} provides two interactions to assemble paragraphs and switch between various contexts generated by LLMs flexibly.

\subsubsection{Quick Assembling}
\begin{figure}[htbp]
    \centering
    \includegraphics[width=\columnwidth]{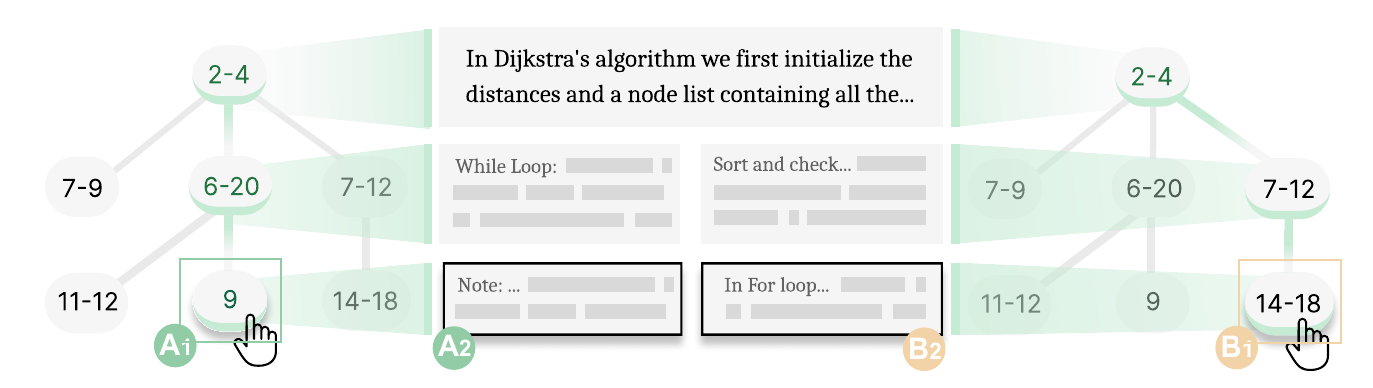}
    \caption{%
    Users can achieve quick assembling nodes by selecting any node in the tree graph as the tail of the chain. For example, after user clicks on node (A1) in outline view, the content in tutorial view (A2) updates correspondingly. (B1) and (B2) is another example.
    }
    \label{fig:int-branch}
\end{figure}

{\tname} supports swift switching between different branches.
When users are not satisfied with the current version of the tutorial and desire to explore additional possibilities, they can select an arbitrary node in the graph as the tail of the chain.
As a result, the selected node and its ancestor nodes will be automatically assembled together and replace the original chain ({\cref{fig:int-branch}}).
By selecting an arbitrary node as the starting point for exploration, users can delve into new directions and uncover untapped potential in the tutorial’s structure and content.
\subsubsection{Step-by-Step Assembling}
Instead of quickly assembling a chain of nodes, {\tname} also provides a deliberate method to manually decide on each step from any selected node in the graph as the entry point.
The branch view (\cref{fig:system}~E) is designed to demonstrate the details of multiple reasoning paths from the selected parent node, which encompasses two aspects:
\vspace{-2pt}
\begin{itemize}[leftmargin=10pt, noitemsep]
    \item \textbf{Choices:}
    When the agent generates several candidate thoughts for the next step, we recommend the best choice through a voting process.
    At each step, the agent casts its vote for the most promising thoughts.
    {\tname} then presents the top 3 choices, visually representing the voting results through the line width that connects the nodes.
    This visual encoding serves as a reference for evaluating the reliability of the diverse thinking paths. 

    \item \textbf{Reasons:}
    {\tname} instructs the agent to provide explicit reasons for its choices.
    Leveraging our prompt strategies, the agent mostly bases its reasoning on the logical connection between code snippets or the coherence of textual paragraphs related to nodes.
    {\tname} presents the textual reason of each choice on the links, enabling users to make more informed and discerning decisions.
\end{itemize}

When a candidate node is chosen, the branch view transitions to a deeper level, allowing users to delve further into exploring the possibilities for the next paragraph.
This process can be repeated iteratively until users finalize the new framework of the tutorial.
Moreover, users retain the capability to navigate back to previous nodes, allowing them to backtrack and revise their decisions whenever necessary.
This iterative and flexible approach empowers users to progressively explore the content of the tutorial.

\subsection{Detail Refinement in Node Space}
\label{sec:system:nodes}
After confirming the structure of the tutorial, users often need to further refine the textual content to align it with their expectations (DR2).
They may not satisfied with the text content pertaining to a particular intent, such as an explanation for specific code fragments, and try to refine it.
To streamline this process, {\tname} implements two-part interactions, providing users with a systematic approach to refine the content and achieve the desired outcome (DR4).



\subsubsection{Exploring Alternatives}
{\tname} offers users the ability to chose alternative representations for a specific intent.
LLM-generated paragraphs are visualized as scattered nodes in the node space view (\cref{fig:system} F1).
We utilize \mbox{OpenAI}'s text embedding model\cite{OpenAI_2023_embed} to measure the semantic similarity of generated paragraphs and projected the embeddings onto a 2D plane using the UMAP algorithm\cite{mcinnes_2018_umap}.
As a result, paragraphs with similar meanings are positioned closer to each other in this 2D space,
which facilitates the quick retrieval of semantically relevant alternatives.

The node alternatives that share the same intent as the user-selected node are highlighted, with color distinguishing their origin, aligning with the color encoding in the outline view (\cref{fig:system}~D).
Nodes are considered to have the same intent if they are generated through the same action type and are associated with the same code fragments. 
For example, two nodes may both describe the first line of code but with distinct representations.
This view enables users to filter and select content that aligns with their specific intent, serving as a foundation for subsequent steps.

\subsubsection{Rewriting Details}
\begin{figure}[htbp]
    \centering
    \includegraphics[width=\columnwidth]{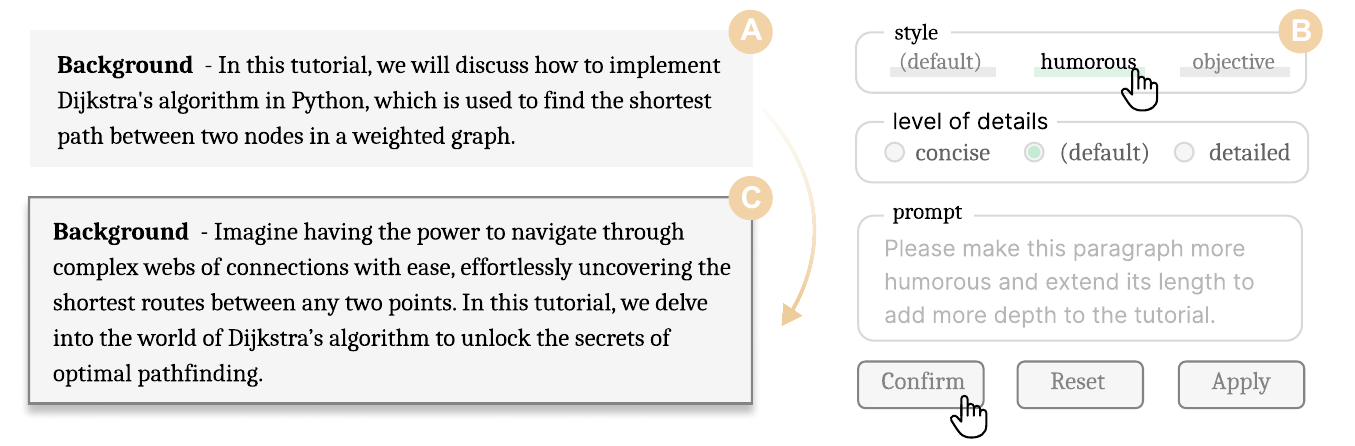}
    \caption{%
    In {\tname}, users can polish the content and adjust the levels of detail of any paragraph (A) by setting options in configuration panel (B), then retrieve polished paragraph (C) from LLM. 
    }
    \label{fig:int-node}
\end{figure}

According to the feedback from the formative study, we conclude several aspects that users primarily focus on when they intend to refine the generated tutorial content.
{\tname} then allows users to customize the content from the following aspects:
\vspace{-2pt}
\begin{itemize}[leftmargin=10pt, noitemsep]
    \item \textbf{Writing styles.}
    Users may have their preferences for the style of the content.
    {\tname} provides several options to make LLMs rephrase the content in an designated style.
    
    \item \textbf{Level of details.} 
    Users might be troubled by the unpredictable initial quantity of LLM-generated content, while an excellent tutorial should be reasonably detailed. 
     Therefore, {\tname} allows users to adjust the level of details while preserving the originally conveyed information.
     \added{\item \minor{\textbf{Prompt Refinement.}} 
     In order to meet a wider range of user needs, in addition to the two aspects mentioned above, {\tname} also supports authors to freely modify the prompt in order to improve the content.}
\end{itemize}
\vspace{-2pt}
As shown in \cref{fig:int-node}, when users find the background is too brief and not attracting enough, they can adjust the level of details and choose a polishing style in the configuration panel, then navigate to the \textit{Confirm} option to prompt the LLM to rephrase the content accordingly.

\subsection{Visualization for Tutorial Understanding}
\label{sec:system:vis}
 To strike a balance between clearly separating two kinds of media resources (i.e., code and text) and offering visual elements of their connections (DR3), {\tname} designs visual representations for users to understand what the text block is about and where it comes from.

\textbf{Brief Block.}
Although the chain view provides the structure of a tutorial explicitly, it is still arduous work for users to grasp the main idea of each paragraph.
Therefore, {\tname} prompts LLMs to brief each paragraph in few words to enable users to quickly comprehend the key concept it intends to convey. 

\textbf{Elements Matching.}
To reduce the switch of users' attention between the code view and the tutorial content, {\tname} presents a set of visual cues for element matching between the source code and the tutorial document.
We extract the connection between the code snippet and paragraph, then present the information in the node.
Furthermore, {\tname} provides visual links of the connection between code snippet and text blocks when users focus on certain node or text block; thus, they can locate the text and code quickly.
During the tutorial generation stage, the focused node will be automatically updated as the latest generated node, enabling users to follow the LLM's generation process more easily and absorb the newly-added information in a more moderate way.

%% file: chapters/6-eval.tex
\section{Technical Evaluation}
To validate the effectiveness of our prompting technique, we conducted a technical evaluation focusing on the accuracy of the extracted text-code connections.

\subsection{Experiment Settings}
\textbf{Dataset.}
We collected 10 code snippets with diverse programming languages, lengths, and structures, covering different categories. 
These snippets were fed into our system to automatically generate programming tutorials.
Then, we manually labeled the code segments that each paragraph references, establishing a ground truth for evaluating text-code connection extraction accuracy. 

\textbf{Procedure.}
\minor{The labeling work and accuracy calculation were conducted by 3 experienced programmers. They separately manually labeled the connection between each paragraph and code segments. They then compared the three labeled results and performed a secondary verification if there were differences among these results. This procedure was designed to guarantee the accuracy of the result, preventing potential errors that may occur during an individual’s labeling process.} After that, they examined the text-code connections extracted by {\tname} to calculate their accuracy.
Subsequent analyses were conducted on failure cases and contributing factors to these errors.

\subsection{Results}

The dataset comprised 10 programming tutorials, encompassing 138 paragraphs and the corresponding text-code connections. 
\added{For the details of the procedure, generated content, and statistical analysis, please refer to the supplemental materials.}

\textbf{Metrics.} The extracted text-code connections exhibited an overall accuracy of \textbf{86.2\%}, comprising 119 correct connections and 19 erroneous ones.

\textbf{Error Analysis.} Of the 19 erroneous connections identified, we categorized them into three types: (1) \emph{No Code} (5/19), (2) \emph{Incorrect Code Range} (12/19), and (3) \emph{Incorrect Code Content} (2/19). For the first error type, the LLM failed to provide any code for the associated paragraph. For the second error type, the LLM provided an inappropriate code range mismatched with the corresponding paragraph. Notably, the majority (10/12) presented the entire code range excessively, whereas only two instances demonstrated incomplete or unrelated connections. For the third error type, the LLM provided fictitious code content misaligned with the original code snippet. We discern these errors as manifestations of LLMs' \emph{hallucination}, particularly evident when employing complex prompting strategies like Tree-of-Thoughts prompting strategy, coupled with a high temperature which might lead to more unexpected results.

\textbf{Summary.} Despite the few failure cases, the results indicate the reliability and accuracy of our prompting technique in extracting text-code connections. This efficiently facilitates users in obtaining faithful content, laying a robust foundation for further refinements of the content.





%% file: chapters/7-user.tex
\section{User Study}

To evaluate the effectiveness of our system in facilitating the programming tutorial authoring process, we conducted a user evaluation to collect feedback on (1) the features of {\tname},
and (2) its support for tutorial authoring compared to a baseline interface. 
\added{Subsequently, we analyzed user interaction logs to gain an understanding of the emergent workflow patterns and user behaviors using {\tname}.}

\subsection{Methodology}
\subsubsection{Participants}
We recruited 12 participants (P1-P12) for our experiment, consisting of 5 females and 7 males, aged 22-28, from a local university.
P1-6 were experienced tutorial authors who had also participated in the formative study (E1-6).
The remaining participants were students with experience in code documentation, but none had experience in authoring programming tutorials.
All participants had more than 4 years of programming experience and had read programming tutorials in the past.
Ten of them were familiar with ChatGPT and regularly utilized it for writing tasks (\emph{e.g.}, ideation, drafting, polishing), while two had heard of it but seldom incorporated it into daily work.
Participants attended our experiment through online meetings, used our system deployed online, and the process was recorded.

\subsubsection{Experiment Condition}

The experiment was conducted under two conditions:
\begin{itemize}
    \item ChatGPT (\emph{baseline}). 
    The baseline interface offered an environment comprising a document editor, a code viewer, and ChatGPT. Users could freely arrange these elements within the interface.
    \item {\tname} (\emph{ours}).
    The system integrated a code viewer and document editor, enriched with a series of interactive visualizations of the creation process with LLMs.
\end{itemize}
Each participant was tasked with creating tutorials using both of the aforementioned tools.
The order of tool usage was counterbalanced among participants: half began with ChatGPT, while the other half started with {\tname}.

\subsubsection{Procedure}
The experiment consists of the following phases:

\textbf{Introduction and Training.}
First, the purpose of the experiment was presented. 
Participants then signed a consent form and filled out a pre-study questionnaire regarding their background and experience with tutorial authoring.
After that, they were introduced to {\tname}'s features, followed by a demonstration of crafting a complete algorithm tutorial using these features.
We gave the participants adequate time to familiarize themselves with the system and encouraged them to reproduce the tutorial independently. All participants were free to ask any clarifying questions.

\textbf{Targeted Task.}
For this task, participants were provided with a piece of code for reference. 
Their objective was to author a tutorial based on a given framework, which involved two sub-tasks: modifying the content of at least one paragraph and describing the main idea of each paragraph. 
Participants underwent two trials in this phase, creating programming tutorials in each trial using two different tools – \emph{baseline} and \emph{ours} – respectively.


\textbf{Open-ended Task.}
For this task, only the problem's background and the source code were provided.
Participants were encouraged to create a step-by-step programming tutorial freely using {\tname}.
They were expected to produce at least three tutorial versions and select one as the final result.
We ensured participants had enough time to validate the generated content.
There were no time constraints on the writing process during trials, allowing them to create tutorials to the best of their abilities.

\textbf{Post-study Interview.}
Upon completion of all trials, we conducted a semi-structured interview with each participant.
They were first invited to fill out a questionnaire with 5-point Likert-scale questions, regarding five features in {\tname}, the general usability, and the support for programming tutorial authoring with our system and the baseline.
Then, we inquired their thoughts on how {\tname} could help their actual workflow, the advantages and limitations of the current system design, as well as their suggestions for potential improvements.

\begin{figure}[htbp]
    \centering
    \includegraphics[width=\columnwidth]{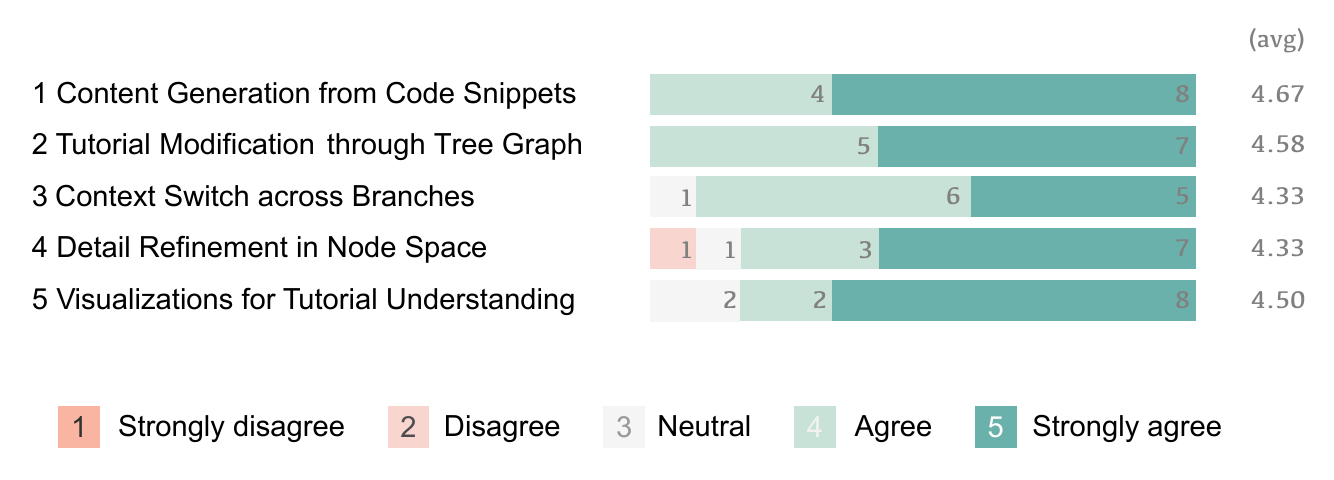}
    \caption{%
    User feedback regarding the features in {\tname}, measured on a 5-point Likert-scale.%
    }
    \label{fig:eval-user}
\end{figure}

\subsection{Results Analysis}
\subsubsection{Features}
We evaluated five features of {\tname} through 5-point Likert-scale questionnaires in the post-study interview.
{\Cref{fig:eval-user}} illustrates the detailed ratings.

\textbf{Participants chose different generation methods based on their needs}.
They responded favorably to the feature of generating content based on selected code, finding it useful when they \textit{``already have an initial framework in mind''} (P6) for authoring content related to familiar code. 
The free generation was also praised since it could \textit{``provide diverse perspectives as a starting point when at lost''} (P3) and allowed them to \textit{``pick up a best one among different versions''} (P1).

\textbf{Participants utilized visual connections to quickly grasp the tutorial content.}
The visual representations of connections were favored by all participants, as it alleviated the effort to \textit{``maintain the connections in mind''} (P5). In traditional chat-based interfaces, even if referred code is provided, participants sometimes still needed to \textit{``scroll back to check the position of the code snippet within the complete code''} (P11) and \textit{``the indentation of the code is missing''} (P2).

\textbf{Participants switched between thoughts smoothly and sensibly.}
According to some participants, utilizing the branch view with thoughts is like \textit{``building text blocks under the guidance of LLMs''} (P3).
Moreover, the quick assembling in Outline by selecting leaf nodes enables \textit{``swift comparison''} (P9) between documents based on different thoughts.

\textbf{Participants employed the tree graph to modify the tutorial.}
Most participants praised the graph-level modifications such as splitting node, describing them as \textit{``novel and convenient interactions within a document''} (P7) that would be \textit{``frequently used''} (P10). For example, P6 split the node which described the main loop in Dijkstra algorithm, because he found the content was too brief. Similarly, P8 utilized ``Trim'' to delete the content describing the engineering process, which \textit{``can be omitted in a tutorial''}.

\textbf{Participants achieved easy detail refinement in node space.}
When focusing on specific content from LLMs, the node space serves as \textit{``a good container which collects all the content for me''} (P4). {\tname} also enables purpose-specific content refinement.
For instance, P9 easily got a concise paragraph in {\tname}, whereas with ChatGPT, he had to test prompts iteratively.
However, P7 expressed dissatisfaction about this feature as his demands were not covered currently, such as introducing metaphors in the tutorial.

\begin{figure}[htbp]
    \centering
    \includegraphics[width=\columnwidth]{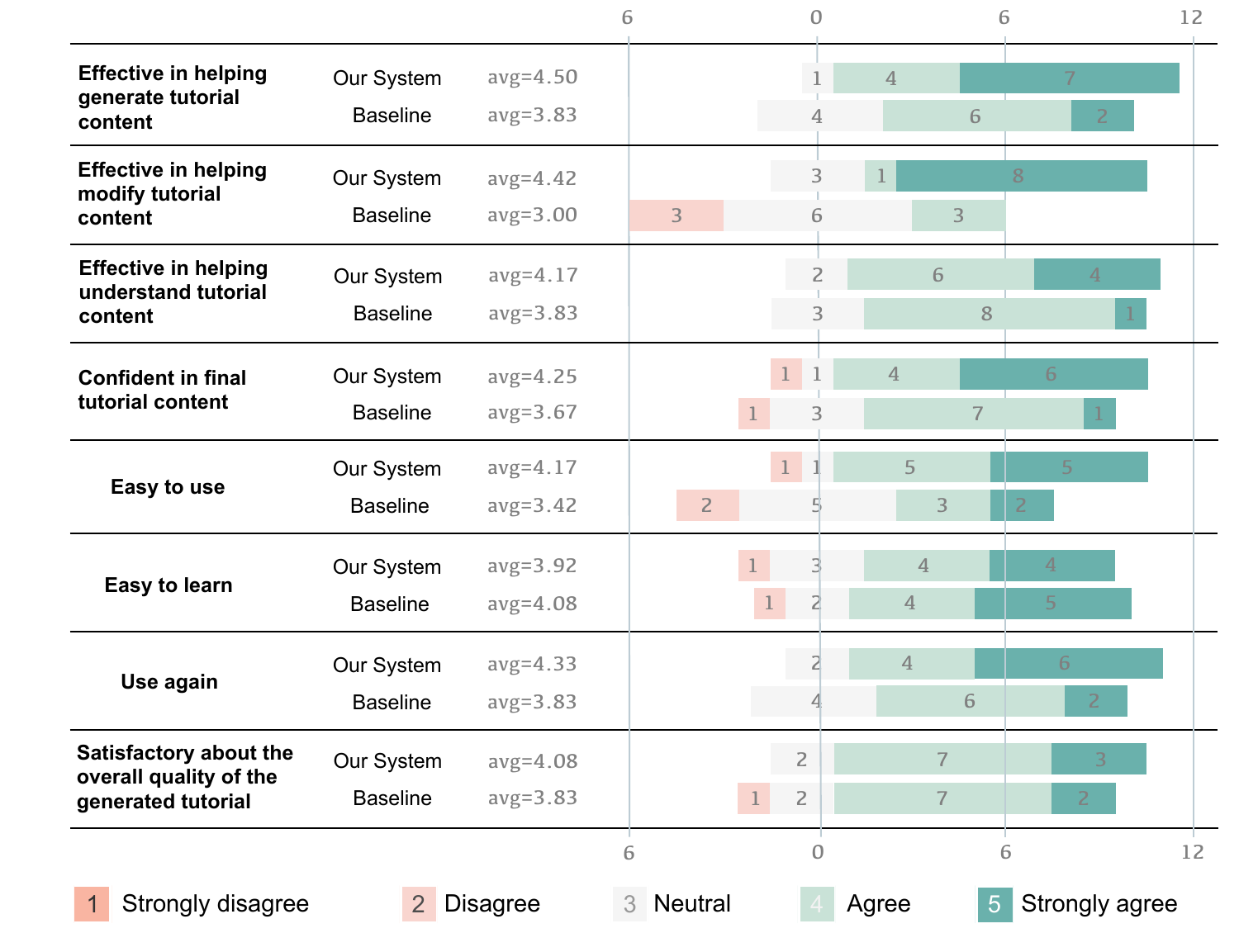}
    \caption{%
    The results of the questionnaire regarding the authoring support
of our system and the baseline.%
    }
    \label{fig:eval-compare}
\end{figure}

\subsubsection{Tutorial Authoring Support Analysis}
At the end of the study, we asked participants to rate the two authoring tools across seven dimensions (shown in {\cref{fig:eval-compare}}). 
Their feedback demonstrated the system usability and its effectiveness in supporting tutorial creation process.

\textbf{{\tname} enables flexible tutorial generation for enhanced engagement.}
Based on participants' feedback, the generation methods in {\tname} allow them to conveniently tailor LLM-generated content to match their expectations \added{($\mu=4.5> 3.83$, $p =.021$)}, offering different degrees of autonomy, while it is challenging with ChatGPT to maintain such control.
\added{For example, P4 started the authoring process with the free generation as the \textit{``initial execution for inspirations''}, paused the process, and then searched for \textit{``a reasonably detailed content from all branches''} in the branch view to serve as a foundation for the tutorial.}
After spotting a favorable structure, he proceeded to manually select the code pieces for content generation, thereby ensuring the resulting output met his expectations.

\textbf{{\tname} provides easier modifications on tutorials hierarchically.}
Participants reported that the interactions provided in {\tname} enabled easy tutorial modifications across various aspects \added{($\mu=4.42>3$, $p = .007$)}, eliminating the need for \textit{``repeatedly writing prompts''} (P11).
Some participants concentrated on the narrative order of the tutorial, while others tended to manipulate the style or detail levels of certain paragraphs. Often, participants employed a combination of these methods. For example, P3 first decided the tutorial structure according to LLM recommendations, then grouped some nodes to shorten the tutorial, and finally refined the text of an unsatisfactory paragraph.
In contrast, when working with ChatGPT, participants had to interact with it in multiple rounds, as the prompts were \textit{``ambiguous to convey their demands in mind''} (P7).

\textbf{{\tname} enhances tutorial comprehension and reliability during the authoring process.}
Participants reflected that they could produce tutorials with more comprehension \added{($\mu=4.17>3.83$, $p=.157$)} and confidence \added{($\mu=4.25>3.67$, $p =.07$)} on the outcomes throughout the workflow of {\tname}.
During the generation process, the connections are updated synchronously, which \textit{``offers clear visual cues''} (P1) about the content of each paragraph.
P12 also pointed out that when utilizing ChatGPT to modify content, the revised content often appeared separately from the original tutorial and code, leaving him uncertain about its correctness.
Moreover, we found that most participants would browse the entire tutorial to guarantee the key points were covered. In {\tname}, the visual connections can support users to \textit{``easily verify the coverage of the provided code''} (P7), thus speeding up the examination process. 

\added{
\textbf{{\tname} elevates the level of engagement and ease for authors.}
Participants appreciated that {\tname} provided a more user-friendly experience in the process of authoring tutorials comparing to ChatGPT ($\mu=4.17> 3.42$, $p=.038$),
without significantly increasing the learning effort ($\mu=3.92<4.08$, $p=.564$).
P2 said the novel visual interface made him \emph{``more engaged''} with the authoring work.
Participants also conveyed a positive inclination towards integrating SPROUT into their real work practices ($\mu=4.33>3.83$, $p=.083$).
}

\added{
\textbf{{\tname} produces user-satisfying tutorials of superior quality.}
Participants expressed their satisfaction with the quality of tutorials generated by {\tname} ($\mu=4.08>3.83$, $p=0.317$), appreciating aspects such as structure, clarity, and educational value.
P5 suggested that our system has the potential to serve as an \emph{``interesting learning tool''}, in addition to its capabilities as an authoring tool.
Beyond the quality of the textual content, the visual design of the system also contributes substantial educational value.
As a teacher, P6 valued the ease with which he could customize tutorials to cater to learners of varying levels with minimal effort.
}

\subsubsection{Authoring Workflow Analysis}
\added{We analyzed the interaction logs from participants to investigate the impact of {\tname} on the authoring workflow.
Our observations revealed diverse workflow patterns adopted by participants in completing their tutorials, along with notable behaviors when interacting with {\tname}.}

\added{
\textbf{Workflow Patterns.}
Two predominant workflow patterns were observed among participants. The most prevalent one was to utilize the automatic generation feature first to produce a complete tutorial, with a primary focus on the tutorial view and chain view. Following the completion of the generation process, participants proceeded to modify the tutorial using the Outline, Branches, and Node Space to suit their specific requirements.
The second pattern emerged when participants desired to take the initiative. In this situation they used user-defined generation, modified the content then moved to the next step, where they switched attention between different views.
Our system offered robust support for both of the aforementioned workflows.

\textbf{Observed Behaviors.} Participants sometimes tended to utilize the features of {\tname} with novel target according to their original authoring habits, which were not considered before. For instance, the demonstration of the reasons in branch view also \minor{assisted} participants to verify the reasoning of tutorials upon completion and during the final verification process.
In terms of tree graph, some experienced participants utilized it to compare different branches for optimal structuring apart from understanding the thinking steps of LLM.
The node space view also served as an organizational tool for participants to efficiently access their desired content among multiple versions of paragraphs resulting from iterative content refinement.

}

%% file: chapters/8-fw.tex
\section{Discussion and Future Work}
In this section, we reflect on our research, mainly discussing the implications, limitations, and future work.

\subsection{Implications}
\textbf{Incorporating LLMs into programming tutorial authoring workflow.}
The development of large language models opens vast possibilities to facilitate the programming tutorial authoring process through reducing manual effort on writing and editing.
While our work provides a step-stone towards programming tutorial authoring with the assistance of LLMs, future work could investigate on further enriching tutorials with more multifaceted content, such as execution examples, flow charts, or figures illustrating code execution processes, which calls for utilizing multi-modal approaches and specific interactions to combine versatile media into current workflow.
Recent state-of-the-art research leveraging in-context learning technologies has been successful in producing code-related content of remarkable quality\cite{ahmed_2023_improving,geng_2024_large}.
Our work, however, concentrates on enhancing the interaction between humans and AI.
Integrating these two approaches with advanced techniques, such as active learning, has the potential to yield more user-centric and efficient tools\minor{\cite{li_2023_human}}.
We believe that our work can be a solid foundation for future researches on authoring programming tutorials.
By incorporating a wider array of materials, we can ultimately create content with higher quality, granting readers a richer understanding of the programming concepts presented.

\textbf{Step-by-step exploratory process offers a more controlled authoring experience.}
Numerous studies have delved into advancing strategies that prompt LLMs to think step by step\cite{Yao_2023_ToT, Yao_2023_ReAct, Wei_2022_CoT} for higher quality content.
Beyond this, we see potential in using this decomposed process as an entry point to grant users broader access to the intermediate steps of the authoring workflow. This empowers them to achieve anticipated outcomes, transforming a typically passive experience into an active one.
The evaluation results indicated that the interactive, segmented representations of the creation process and generated tutorial content in {\tname} offer more precise generation and allow for multi-level adjustments to modules.
Such an approach addresses the often-encountered challenge of accurately conveying intentions in chat-based interfaces.


\textbf{Maintaining connections between provided inputs and LLM-generated outputs ensures transparency and traceability.}
We identify and visualize latent connections between the source code and the generated tutorial, as we believe it can assist users to comprehend and verify the reasoning behind LLMs' creation process.
Informed by our evaluation results, users utilized these presented connections to examine the outcomes after generating or modifying procedures, easing the burden of spending extra cognitive efforts to recognize and memorize them.
Moreover, consistently maintaining these connections significantly improves user confidence in the final results crafted collaboratively by users and LLMs.
This is essential in materials-backed authoring scenarios, where consistency and correctness are vital.
Besides, our LLM-based methodology and visual design strategies for establishing connection between cross-modal data promise to be beneficial across a wide range of multi-modal creation and data analysis scenarios\cite{feng_2022_ipoet, feng_2023_xnli,Wang_2023_ai}.


\textbf{Striking a balance on the degree of interaction integrity in interface design.}
A key finding is that user preferences for interacting with LLMs vary by individuals.
For example, depending on the specific operations they intend to perform, some individuals preferred utilizing integrated interactions as a means to effortlessly obtain accurate outcomes, while others expressed reservations regarding the comprehensiveness of these approaches in covering all possible scenarios.
There exists a conflict between the extensiveness of features coverage and the precision of the results after execution when designing interactions.
Thus, our research could inspire future studies to more systematically explore the interface design paradigm for LLM-based applications, which could range from completely-free form to structured module-based interactions.

\added{\textbf{Applying methodologies to diverse LLM-based creation scenarios.}
User feedback from our evaluation underscores the importance of both fidelity and innovative capabilities of LLMs in authoring tasks.
Designing and developing systems or tools for creative scenarios requires a harmonious mix of adaptability and innovation to ensure their use is seamless and user-friendly\cite{fan_2023_largesurvey}.
Although our study focuses on utilizing novel prompt strategies and interactive visualizations for programming tutorial authoring, the methodologies employed herein can potentially be extended to other creation scenarios that require consistency with reference materials, such as fact-based analysis for legal documents\cite{cui_2023_chatlaw}, medical applications\cite{thirunavukarasu_2023_medicine, Jin_2023_medi}.
The underlying idea involves designing purpose-specific and scenario-oriented prompting methods to enable appropriate exploratory workflows.
\minor{The visualization of the generation process of LLMs can also be generalized to the majority of authoring tasks\cite{sultanum_2023_datatales,Zhang2023VISARAH,yuan_2022_wordcraft} to make the process more tracable.}
Furthermore, the value of a step-by-step rationale extends beyond authoring application, with potential benefits to other domains such as commonsense Q\&A \cite{ocker_2023_comm} or text learning\cite{kasneci_2023_education}.
By breaking down complex or lengthy content into manageable segments, the learning process becomes more palatable through this incremental methodology. 
This suggests the possibility of integrating the step-by-step methodology into applications aimed at enhancing the learning workflow (e.g., mathematical concepts and language learning), which can offer a more structured and digestible approach to knowledge acquisition\cite{Yu_2023_pub}.}

\subsection{Limitations and Future Work}
\label{sec:limitation}
Although our work has shown promising results, there are several limitations and opportunities for future research.

\textbf{Scalability.}
Although our system effectively supports tutorial authoring process in most cases, it may face limitations in adequately covering the context when dealing with lengthy source code and tutorials, primarily due to token limitations.
Nevertheless, these limitations can be addressed with advancements in LLM capabilities or by utilizing memory summarization techniques.

\textbf{Potentiality.}
While our study focuses on modifying content within a single chain of thought, the current design framework can support more varied tutorial revision operations.
For example, future work could explore facilitating users in merging thoughts from two distinct paths, ensuring context preservation. This enhancement, rooted in our current prompting strategies, would also necessitate broadening the scope of graph manipulations to enable more flexible interactions for users to customize the content.

\textbf{Generalizability.}
Further exploration into multi-modal approaches can benefit tasks involving non-textual elements, like data visualization.
Based on the nature of source materials and expected outcomes, visual representations beyond tree graphs should be further explored to enhance the comprehension and manipulation during the creation process.
For example, future researches can investigate how to properly reify the creation process with LLMs when handling non-linear structured data.

\textbf{Limitations.}
While the current prompting methods are sufficiently reliable and accurate for the authoring workflow, the extraction accuracy can be further enhanced.
Future work could focus on utilizing correction prompts to examine the extracted connections or integrating add-on modules to identify connections to guarantee a higher accuracy.
\added{Additionally, while insightful, our study is limited by its laboratory setting as it does not encompass assessments in real educational scenarios.
Introducing student agents for tutorial assessment and iterative modification could potentially enhance its education value.
Conducting a evaluation set in real education scenarios is also suggested for a more holistic understanding of the realistic learning impact.}


%% file: chapters/9-con.tex
\section{Conclusion}

This work presents {\tname}, a programming tutorial authoring tool with LLMs which breaks down the programming tutorial authoring task into actionable steps and adopts novel prompting strategies to generate high-quality and diverse tutorial content. With a series of interactive visualizations of the LLM generation process, users are able to effectively engage in an exploratory process to generate, modify, and understand the generated tutorial.
A user study with 12 participants demonstrates that users can effectively utilize {\tname} to author programming tutorials with LLMs in a transparent and traceable process, leading to more controllable and reliable results.